\documentclass[pdflatex,sn-mathphys,Numbered,12pt]{sn-jnl}
\usepackage{graphicx}%
\usepackage{multirow}%
\usepackage{amsmath,amssymb,amsfonts}%
\usepackage{amsthm}%
\usepackage{mathrsfs}%
\usepackage[title]{appendix}%
\usepackage{xcolor}%
\usepackage{textcomp}%
\usepackage{manyfoot}%
\usepackage{booktabs}%
\usepackage{algorithm}%
\usepackage{algorithmicx}%
\usepackage{algpseudocode}%
\usepackage{cancel}
\usepackage{listings}

\begin{document}

\title[]{Irregular gyration of a two-dimensional random-acceleration process in a confining potential}

\author[1]{\fnm{Victor} \sur{Dotsenko}}\email{dotsenko@lptmc.jussieu.fr}

\author[1]{\fnm{Gleb} \sur{Oshanin}}\email{gleb.oshanin@sorbonne-universite.fr}

\author[2,3]{\fnm{Leonid} \sur{Pastur}}\email{pastur2001@yahoo.com}

\author*[1]{\fnm{Pascal} \sur{Viot}}\email{viot@lptmc.jussieu.fr}

\affil*[1]{\orgdiv{Laboratoire de Physique  Th\'eorique de la Mati\`ere Condens\'ee (UMR CNRS 7600)}, \orgname{Sorbonne Universit\'e}, \orgaddress{\street{4 place Jussieu}, \city{Paris}, \postcode{75005}, \country{France}}}

\affil[2]{\orgname{B. Verkin Institute for Low Temperature Physics and Engineering}, \orgaddress{\street{47 Nauky Ave.}, \city{Kharkiv}, \postcode{61103}, \country{Ukraine}}}

\affil[3]{\orgdiv{King's College London},
\orgaddress{\street{ Strand}, \city{London}, \postcode{WC2R 2LS}, \country{United Kingdom}}}

\abstract{ We study the stochastic dynamics of a two-dimensional particle assuming that the components of its position
are two coupled random-acceleration processes evolving in a confining parabolic potential and are the subjects of  independent Gaussian white noises with different amplitudes (temperatures).  We determine the standard characteristic properties, i.e., the moments of position's components and their velocities, mixed moments and two-time correlations, as well as the position-velocity probability density function (pdf). We  show that if the amplitudes of the noises are not equal, then the particle
experiences a non-zero (on average) torque,  such that
the angular momentum L and the angular velocity W have non-zero mean values. Both are
 (irregularly) oscillating with time t, such that the characteristics of a rotational motion are changing their signs.
We also evaluate the pdf-s of  L and  W and show that the former has exponential tails for any fixed t, and hence, all moments. In addition, in the large-time limit this pdf converges to a uniform distribution with a diverging variance. The pdf of W possesses heavy power-law tails such that the mean W is the only existing moment. This pdf converges to a limiting form which, surprisingly, is completely independent of the amplitudes of noises.
}


\maketitle

\tableofcontents

\vspace{0.2in}

\section{Introduction}

When a physical system is brought in contact with several  heat
baths maintained at different temperatures, the whole
system evolves towards a non-equilibrium steady-state characterized by a
non-zero current and non-equilibrium fluctuations. One-dimensional solvable
models have been developed to demonstrate this behavior \cite%
{Bodineau2004,Visco2006,Fogedby2011} as well as the validity
of various fluctuation theorems and relations (see, e.g., \cite%
{Evans,Marini,Rondoni2007,Seifert}).

In higher dimensions, the Brownian gyrator model (BGM) provides an exactly
solvable, albeit a minimalist example of a system that exhibits such a
non-trivial non-equilibrium behavior. Mathematically, the model is defined
in $d$-dimensions in terms of $d$ linearly coupled Ornstein-Uhlenbeck (OU)
processes (see, e.g., \cite{Wang1945}) each being subject to its own
zero-mean Gaussian white noise, which is statistically-independent of the
noises acting on the other components and is characterized by its own
temperature. In general, the temperatures of the noises are not all equal to
each other.

The BGM was first introduced in \cite{peliti} for the analysis
of the emerging steady-state and its effective temperature in a system of
two coupled single-spin paramagnets each being in contact with its own
thermal bath. Some years after, it was argued in \cite{rei} that, in fact,
the BGM represents a minimal heat engine: there emerges a non-zero current
which moreover has a non-zero curl, generating therefore a non-zero torque.
As a result, if $d$ OU processes are used to describe the components
of an instantaneous position of a particle living in a $d$-dimensional
space, then the particle gyrates around the origin, which
explains the name of the model.

Various aspects of the dynamical and the steady-state behavior of the BGM
have been analyzed in case of standard delta-correlated in time noises
acting on the position components of the particle (see, e.g., \cite%
{cil1,cil2,Berut2014,Dotsenko2013,Fogedby2017,Mancois2018,Bae2021,Alberici2021,NEXUS,Baldassarri2020}%
) and also for the noises with long-ranged temporal correlations \cite%
{Nascimento2021,Squarcini2022a}. Few examples of problems that have been
studied are the entropy production \cite{Crisanti2012,Mazzolo2023}, effects
of anisotropic fluctuations \cite{olga,olga2} and non-harmonic potentials
\cite{Chang2021},  rotational dynamics of a Brownian ellipsoid in isotropic potential or of an inertial chiral ellipsoid  \cite{saha},
the spectral properties of individual trajectories of a gyrator \cite%
{Squarcini2022b,sara}, emerging cooperative behavior of several Brownian
gyrators \cite{Dotsenko2022,Dotsenko2023}, as well as the synchronization of
two out-of-equilibrium Kuramoto oscillators living at different temperatures
\cite{Dotsenko2019}. By exerting an external force on the Brownian gyrator,
it was possible to derive the asymmetry relations obeyed by the position
probability density \cite{Cerasoli2018,Cerasoli2021,cherail}.
The
probability densities of the gyration characteristics -- the angular
momentum and the angular velocity -- have been evaluated in recent work \cite%
{viot2023destructive},  while in \cite{touchette,Buisson2023} the moment-generating function of a time-averaged
angular momentum has been studied within a somewhat different context.

The BGM, being  clearly a minimalist model,  is
nonetheless quite realistic and moreover, is experimentally-realizable. In
\cite{cil1,cil2,Berut2014} the BGM was conceived by constructing an
effective electric circuit comprising a capacitance and two resistors kept
at different temperatures. In turn, in \cite{Argun2017} the BGM was
ingeniously devised very directly in a system with a single colloidal
Brownian particle that is optically trapped in an elliptical potential well
and coupled simultaneously to two heat baths with different temperatures
acting along perpendicular directions. Some additional realizations of the
model have been mentioned in \cite{rei} and \cite%
{Mancois2018,fakhri2,Sou2019}.

Most of available analytical analyses, with an exception of those in \cite%
{Mancois2018,Bae2021}, were concerned with the simplest case of a massless
particle, i.e., the effects of inertia were discarded. While such effects
were only briefly discussed in \cite{Mancois2018}, a more systematic
analysis was developed in \cite{Bae2021} by using an appropriately
generalized Langevin description and numerical simulations. The analysis in
\cite{Bae2021} evidenced some interesting effects of the inertia in the
steady-state such as, e.g., a reduction of the non-equilibrium effects by
diminishing the declination of the probability density and the mean
value of a specific angular momentum. It was also demonstrated that
rotation is maximized at a particular anisotropy while the stability of the
rotation is minimized at a particular anisotropy or mass. We note, however,
that the analyses in \cite{Mancois2018,Bae2021} were focused exclusively on
 the behavior in the steady-state attained in
the limit $t \to \infty$.

In this  paper we study a two-dimensional BGM with inertia and
concentrate on the \textit{temporal} evolution of the characteristic
properties of the gyration process. To emphasize solely the effects of
inertia and to make our analysis more transparent and simple, we set
throughout the paper the damping constant equal to zero. The effects of a
finite friction will be addressed elsewhere. This case  is also of interest from a purely
mathematical perspective -- in this situations one deals with two coupled
so-called random-acceleration processes, each living at its own temperature
and both evolving in a parabolic potential. The random-acceleration process
in one-dimensional systems has received much interest within the last decade
as a simple example of a super-diffusive non-Markovian stochastic process
which exhibits ageing and a non-trivial behavior of the extremal properties
\cite%
{Masoliver1995,Masoliver1996,Bray2013,Burkhardt,Bicout2000,Levernier2018,Capala2021,Basu}%
.

We we will show that, not counter-intuitively, the dynamics of the
system under study demonstrate a completely different behavior as compared to that of the
standard BGM. In the absence of  dissipation, an ongoing pumping of
energy delocalizes the particle such that its mean-squared distance from the
origin grows linearly with time, while for the BGM it tends to a constant as
$t \to \infty$. As a consequence, the steady state does not exist. On
average, the particle still performs a rotational motion (likewise the BGM)
when the temperatures of the components are not equal to each other  and when the confining potential is anisotropic, but it
gradually travels away from the origin. The temporal evolution of the
characteristic properties of the gyration process is also strikingly
different from that of the standard BGM. In particular, the mean angular
moment (which approaches a constant value for the BGM) in our case oscillates with time
between its maximal and minimal  values which have different
signs. Correspondingly, the torque imposed on the particle also oscillates
prompting the particle to rotate clockwise and counter-clockwise at
different time intervals. The time-averaged mean angular momentum approaches
nonetheless a constant value as $t \to \infty$, such that the particle
eventually gyrates around the origin, on average, and the gyration direction
is defined solely by the relation between the temperatures of the
components. Accordingly, the mean angular velocity is an oscillatory
function of time, and the amplitude of oscillations decreases as the first
inverse power of time due to an  displacement of the particle away
from the origin.

Further on, motivated by the analysis in recent work  \cite{viot2023destructive}, we
calculate the full probability densities of the angular momentum and the angular velocity. We find that, at a finite time $t$, the
probability density of the angular momentum has exponential tails
and hence, all moments are finite. However, this distribution is effectively
broad and its variance diverges with $t$ as $t^2$, while the expectation is bounded in $t$. This
signifies that the expectation does not  bear  significant physical information and
only indicates a certain non-zero trend (once the temperatures of the components
are unequal) in an ensemble of the processes under study rather than the
behavior to be observed for each individual realization. Thus, the complete description of
such strongly  fluctuating observables is given only by their probability
distribution. This is, in particular, clear for the observables with infinite second moment. We show that this is the case for the angular velocity whose
probability density has heavy power-law
tails so  that the mean angular velocity is the only finite moment.  Curiously enough,  in the limit $t \to
\infty$, the probability density  of the angular velocity converges
to a remarkably simple limiting form which is completely independent of the
temperatures of the components.
Therefore, one  observes quite a non-trivial irregular behavior of the angular momentum and the angular velocity as
functions of time, with quite significant fluctuations.

 The paper is organized as follows: in Sec.\ref{sec:model}, we formulate the
model and introduce basic notations. In Sec. \ref{sec:3} we concentrate on the moments and the cross-moments of
the particle's position and velocities, while in Sec. \ref{sec:twotime} we discuss the behavior of the
 two-time correlation function of the particle's position. Section \ref{sec:Pt} is devoted to the position-velocity probability densities, and in Sec. \ref{sec:energy} we analyze the statistical properties of the kinetic, potential and total energy.
 Next, in Sec. \ref{sec:gyration} we focus on the gyration characteristics, i.e., the
  angular momentum and the angular velocity. We derive explicit expressions for the mean values of these characteristic properties and also their full probability densities.
 Finally, in Sec. \ref{sec:conc} we conclude with a brief recapitulation of our results and a short discussion.

\section{The model}

\label{sec:model}

We consider the dynamics of a particle of mass $m$ in  a two-dimensional system, in the presence of a
potential that consists of two parts:

\begin{itemize}
\item[(i)] The confining potential which we assume, following the trend of
the field, to be the paraboloid
\begin{equation}
U(x,y)\;=\;\lambda \Bigl(\frac{1}{2}\,x^{2}\;+\;\frac{1}{2}\,y^{2}\;+\;u\,xy%
\Bigr),  \label{3}
\end{equation}%
where $\lambda >0$ controls the amplitude of attraction to the origin (e.g.,
the strength of an optical trap), while $u$ is the (relative to $\lambda $) coupling parameter
between the $x$- and $y$-components.  We assume
that $|u|<1$, to guarantee that $U$ is a paraboloid  and hence, that $%
\exp(-U) $ is integrable.

\item[(ii)] The additive noise potential
\begin{equation}
\Xi (x,y,t)=x\xi _{1}(t)+y\xi _{2}(t),  \label{noise}
\end{equation}%
where $\xi _{1}(t)$ and $\xi _{2}(t)$ are mutually-independent stochastic
noises with zero mean and the covariances
\begin{equation}\label{noises}
\begin{split}
\left\langle \xi _{x}(t^{\prime })\xi _{x}(t)\right\rangle & =2T_{1}\delta
(t-t^{\prime })\,, \\
\left\langle \xi _{y}(t^{\prime })\xi _{y}(t)\right\rangle & =2T_{2}\delta
(t-t^{\prime })\,.
\end{split}%
\end{equation}%
The angle brackets in Eqs.~(\ref{noises}) denote averaging with respect to
realizations of thermal noises, and $T_{1}$ and $T_{2}$ are the respective
"temperatures" of the two thermal baths, (which are in general unequal).
 We will characterize the situation as equilibrium or non-equilibrium in cases where $T_1=T_2$ or $T_1 \neq T_2$ respectively. \end{itemize}
Denoting by $\mathbf{r}(t)=(x(t), \,  y(t))$ the instantaneous coordinates of the
particle, we can write down its equation of motion as \footnote{%
Recall that in the standard BGM one has the first derivatives of the
position components $x(t)$ and $y(t)$ with respect to time, in place of
terms $m\ddot{x}(t)$ and $m\ddot{y}(t)$, i.e;
\begin{equation}
\nu \,\dot{x}(t)+\lambda x(t) + \lambda u y (t)= \xi _{1}(t)\,,\qquad \nu \,%
\dot{y}(t)+ \lambda y(t) +\lambda u x(t) =\xi _{2}(t)\,,  \notag  \label{20}
\\
\end{equation}%
where $\nu $ is the damping constant and the potential is defined in Eq. %
\eqref{3}.} 

\begin{equation}\label{2}
\begin{split}
m\,\ddot{x}(t)+\lambda x(t)+\lambda u y(t)& =\xi_{1}(t)\,, \\
m\,\ddot{y}(t)+\lambda y(t) +\lambda u x(t)& =\xi_{2}(t)\,.
\end{split}%
\end{equation}
To define uniquely the dynamics we have to add the initial conditions. We
will be mostly interested by the large-time dynamics of the particle
due to the presence of the noise,   and hence, it seems quite natural to assume, for simplicity, that the particle starts from the
very bottom of the confining potential. Without any significant lack of generality, we also stipulate that the particle's velocity
is equal zero at time $t =0$. Therefore, Eqs. \eqref{2} are to be solved subject to the initial conditions
\begin{equation}
\begin{split}
\label{initial}
x(0)&=y(0)=0 , \\
\dot{x}(0) &= \dot{y}(0) = 0 \,.
    \end{split}
\end{equation}%
The general case of non-zero initial
conditions can be easily obtained from the case (\ref{initial}) by replacing $%
\mathbf{r}(t)$ above by $\mathbf{r}(t)+\mathbf{r}_{0}(t)$, where $\mathbf{r}%
_{0}(t)$ is the standard linear combination of simple harmonic functions,
the solution of Eqs.~\eqref{2}  with given initial conditions and zero  right-hand-side (r.h.s). \textbf{%
\ }

There are several properties of focal interest, which characterize the
stochastic process defined in Eqs. \eqref{noises} -- \eqref{initial}. First of all,
these are the moments of instantaneous positions $x(t)$ and $y(t)$ and the
velocities $\dot{x}(t)$ and $\dot{y}(t)$, as well as their mixed moments.
More generally, this is the joint position-velocity probability density $%
P_{t}=P_{t}(x,y,\dot{x},\dot{y})$. Taking into account that $\Xi $ of Eqs.~ (\ref%
{noise}) -- (\ref{noises}) is a pair of independent white noises, the
collection

\begin{equation}
X=\{X_{j}\}_{j=1}^{4}=\{x,y,\dot{x},\dot{y}\}  \label{X}
\end{equation}%
forms a four component Markov process, hence, their  $P_{t}$ solves
the corresponding Fokker-Planck equation. This proved to be a fairly efficient and
adequate tool in the case of non-linear dynamics and/or non Gaussian but
Markovian $\Xi $ (see e.g. \cite{Lifshits1988}), where the collection $%
\{X,\Xi \}$ forms a Markov process. In our case of linear dynamics and
Gaussian white noise $\Xi $ of Eqs.~(\ref{noise}) -- (\ref{noises}) the
collection, Eq.~(\ref{X}), is also Gaussian. In addition it follows from Eqs. \eqref%
{noises} to \eqref{initial} that
\begin{equation}\label{mX0}
\langle X_{j}(t)\rangle =0,\;j=1,2,3,4.
\end{equation}%
Thus, it suffices to find the (covariance) matrix of $X(t)$,
\begin{eqnarray}\label{eq:MatM}M(t) &=&\{m_{jk}\}_{j,k=1}^{4},\;m_{jk}=\langle X_{j}(t)X_{k}(t)\rangle ,
\notag \\ 
M(t) &=&\left(
\begin{array}{cccc}
\langle x^{2}(t)\rangle  & \langle x(t)y(t)\rangle  & \langle x(t)\dot{x}%
(t)\rangle  & \langle x(t)\dot{y}(t)\rangle  \\
\langle x(t)y(t)\rangle  & \langle y^{2}(t)\rangle  & \langle y(t)\dot{x}%
(t)\rangle  & \langle y(t)\dot{y}(t)\rangle  \\
\langle x(t)\dot{x}(t)\rangle  & \langle y(t)\dot{x}(t)\rangle  & \langle
\dot{x}^{2}(t)\rangle  & \langle \dot{x}(t)\dot{y}(t)\rangle  \\
\langle x(t)\dot{y}(t)\rangle  & \langle y(t)\dot{y}(t)\rangle  & \langle
\dot{x}(t)\dot{y}(t)\rangle  & \langle \dot{y}^{2}(t)\rangle
\end{array}%
\right)  , 
\end{eqnarray}%
 to be able to obtain its characteristic function
\begin{equation}
\Phi _{t}(\boldsymbol{\omega })=\int_{\mathbb{R}^{4}}e^{i(\boldsymbol{\omega }%
,X)}\,P_{t}(X)dX\,=\exp \{-(M(t)\boldsymbol{\omega },\boldsymbol{\omega })/2\},\;%
\boldsymbol{\omega }=\{{\omega _{j}\}}_{j=1}^{4}.  \label{phit}
\end{equation}
Likewise, the corresponding joint probability distribution density $P_{t}$
of $X(t)$ is, in general
\begin{equation}
P_{t}(X)=((2\pi )^{4}\det M(t))^{-1/2}\exp \{-(M^{-1}(t)X,X)/2\}.
\label{dista}
\end{equation}%
However, for this formula to be well defined, the covariance matrix
(\ref{eq:MatM}) has to admit the inverse matrix $M^{-1}(t)$. It follows from the results of Sections 3.1 - 3.2 that the leading terms of the large-$t$ asymptotics of the diagonal entries of $M(t)$ are proportional to $t$, while the off-diagonal entries  are bounded in $t$.  This implies the invertibility of $M(t)$ if $t$ is large enough. A more sophisticated argument shows that $M(t)$ is invertible for any $t>0$ (see Appendix \ref{secA0}).

The knowledge of $\Phi _{t}$ or/and $P_{t}$ allow one to study the behavior of two
important quantities that characterize, as in the case of the BGM, the emerging rotational motion of the particle.

The first is its angular momentum%
\begin{equation}
\mathbf{L}(t)=m(\mathbf{r}(t)\times \mathbf{v}(t))\,,\; \mathbf{v}(t)=\dot{\mathbf{r}}(t), \label{meanLLL}
\end{equation}%
where $\mathbf{r}(t)\times \mathbf{v}(t)$ is the cross product of the
position and the velocity of the particle, so that $\mathbf{L}$ is
perpendicular to the plane of motion
\begin{equation}
\mathbf{L}(t)=(0,0,L),\;L(t)=m\left( x(t)\dot{y}(t)-y(t)\dot{x}(t)\right) \,.
\label{meanL}
\end{equation}%
The second characteristic of interest is the angular velocity, which is
formally defined as
\begin{equation}
\mathbf{W}(t)=(0,0,W(t)),\;W(t)=\,\frac{x(t)\dot{y}(t)-y(t)\dot{x}(t)}{%
x^{2}(t)+y^{2}(t)}\,.  \label{meanomega}
\end{equation}%
We will study below certain moments of these random variables, and will also
derive exact expressions for their probability densities.

Lastly, to simplify the subsequent formulas, we set $m=1$ in what follows.
The dependence of $m$ can be easily recovered in our final results by simply
multiplying $L(t)$ and $W(t)$ by $m$, by replacing $\lambda $ by $\lambda /m$, and
the temperatures $T_{1,2}$ by $T_{1,2}/m^{2}$.

\section{Particle's trajectories and velocities, and their moments}

\label{sec:3}

We will omit the argument $t$ in many functions below where this does not lead to a confusion.

The solution of Eqs. \eqref{2} and \eqref{initial} for a fixed realization of noises, Eqs. \eqref{noise} and \eqref{noises},
is
\begin{equation}\label{a}
\begin{split}
x=x(t)& =\frac{1}{2}\int_{0}^{t}d\tau \Bigg[Q_{+}(t-\tau )\xi _{1}(\tau
)+Q_{-}(t-\tau )\xi _{2}(\tau )\Bigg]\,, \\
y=y(t)& =\frac{1}{2}\int_{0}^{t}d\tau \Bigg[Q_{-}(t-\tau )\xi _{1}(\tau
)+Q_{+}(t-\tau )\xi _{2}(\tau )\Bigg]\,,
\end{split}%
\end{equation}%
where we denoted
\begin{equation}\label{b}
Q_{\pm }(t)=\frac{\sin \left( \Omega _{+}t\right) }{\Omega _{+}}\pm \frac{%
\sin \left( \Omega _{-}t\right) }{\Omega _{-}}\,,\quad \Omega _{\pm }=\sqrt{%
\lambda (1\pm u)}\,,\,\,|u|<1\,.
\end{equation}%
In turn, from Eqs. \eqref{a} and \eqref{b} we find the instantaneous
velocities $\dot{x}(t)$ and $\dot{y}(t)$ :
\begin{equation}
\begin{split}
\label{c}
\dot{x}=\dot{x}(t)& =\frac{1}{2}\int_{0}^{t}d\tau \Bigg[\dot{Q}_{+}(t-\tau
)\xi _{1}(\tau )+\dot{Q}_{-}(t-\tau )\xi _{2}(\tau )\Bigg]\,, \\
\dot{y}=\dot{y}(t)& =\frac{1}{2}\int_{0}^{t}d\tau \Bigg[\dot{Q}_{-}(t-\tau
)\xi _{1}(\tau )+\dot{Q}_{+}(t-\tau )\xi _{2}(\tau )\Bigg]\,,
\end{split}%
\end{equation}%
with
\begin{equation}\label{d}
\dot{Q}_{\pm }(t)=\cos \left( \Omega _{+}t\right) \pm \cos \left( \Omega
_{-}t\right) \,.
\end{equation}%

\smallskip
We will analyze now the behavior of the moments of the instantaneous
positions and velocities. 

\subsection{Second moments of positions}

Explicit expressions for the mean-squared displacements from the origin
follow readily from Eqs. \eqref{a} and \eqref{b} :
\begin{equation}  \label{x}
\begin{split}
\left \langle x^2 \right \rangle &= \frac{\left(T_1 + T_2\right)}{ 2 \lambda
(1 - u^2)} \, t - \frac{\left(T_1 + T_2\right)}{8} \left(\frac{\sin\left(2
\Omega_{+} \, t\right)}{\Omega_+^{3}} + \frac{\sin\left(2 \Omega_- \,
t\right)}{\Omega_-^{3}} \right) \\
&+ \frac{\left(T_1 - T_2\right)}{2 u \lambda } \left[\frac{%
\cos\left(\Omega_- \, t\right)\sin\left(\Omega_+ \, t\right)}{\Omega_+} -
\frac{\cos\left(\Omega_+ \, t\right)\sin\left(\Omega_- \, t\right)}{\Omega_-}
\right] \,,
\end{split}%
\end{equation}
and
\begin{equation}  \label{y}
\begin{split}
\left \langle y^2 \right \rangle &= \frac{\left(T_1 + T_2\right)}{ 2 \lambda
(1 - u^2)} \, t - \frac{\left(T_1 + T_2\right)}{8 } \left(\frac{\sin\left(2
\Omega_+\, t\right)}{\Omega_+^{3}} + \frac{\sin\left(2 \Omega_- \, t\right)}{%
\Omega_-^{3}} \right) \\
&- \frac{\left(T_1 - T_2\right)}{2 u \lambda} \left[\frac{\cos\left(\Omega_-
\, t\right)\sin\left(\Omega_+ \, t\right)}{\Omega_+} - \frac{%
\cos\left(\Omega_+ \, t\right)\sin\left(\Omega_- \, t\right)}{\Omega_-} %
\right] \,.
\end{split}%
\end{equation}

We observe that $\left\langle x^{2}\right\rangle$  and $\left\langle
y^{2}\right\rangle$  differ only if $T_{1}\neq T_{2}$ ("non-equilibrium"
situation) as they should, due to the terms in the second lines in the
l.h.s. of Eqs. \eqref{x} and \eqref{y}. These terms, proportional to
$T_{1}-T_{2}$, are denoted here and below by square brackets. We note that
these terms are bounded as $t\rightarrow \infty $, and, hence, determine only
a transient behavior. The leading large-$t$ behavior of $\langle
x^{2}\rangle $ and $\langle y^{2}\rangle $ is identical and comes from the
first terms in Eqs. \eqref{x} and \eqref{y} which grow linearly with $t$. In
other words, the particle becomes delocalized (as compared to the standard
BGM) and travels away from the origin despite the confining potential. Note also that the frequencies $\Omega_{pm}$ are continuous functions of $\lambda$ and $u$, hence, are incommensurate generically, resulting in the irregular oscillations of all terms except the first ones in Eqs. \eqref{x} and \eqref{y}.

It is also worth mentioning that in the limit $\lambda
\rightarrow 0$, in which the potential in Eq. \eqref{3} vanishes, Eqs. %
\eqref{x} and \eqref{y} predict a much faster growth
\begin{equation}
\langle x^{2}\rangle =\frac{2T_{1}}{3}\,t^{3}\,,\quad \langle y^{2}\rangle =%
\frac{2T_{2}}{3}\,t^{3}\,,\quad \lambda =0,
\label{e}
\end{equation}%
which is a well-known result for the random-acceleration process \cite%
{Masoliver1995,Masoliver1996,Bray2013,Burkhardt,Bicout2000,Levernier2018,Capala2021}
(see also recent \cite{Basu} and references therein). In addition, for $u=0$%
, when the coordinates decouple, while $\lambda >0$, we have two independent
random-acceleration processes evolving in a quadratic potential. In this
case, we get from Eqs. \eqref{x} and \eqref{y}
\begin{equation}
\begin{split}
\langle x^{2}\rangle & =\frac{T_{1}}{\lambda }\,t-\frac{T_{1}}{2\lambda
^{3/2}}\sin \left( 2\sqrt{\lambda }t\right) \,, \\
\ \langle y^{2}\rangle & =\frac{T_{2}}{\lambda }\,t-\frac{T_{2}}{2\lambda
^{3/2}}\sin \left( 2\sqrt{\lambda }t\right) \,,\quad u=0\,,\quad \lambda
>0\,,
\end{split}%
\end{equation}%
i.e., that the mean square displacements increase linearly with time and
have additional oscillating terms, whose frequency is determined by the
amplitude of the potential.

Therefore, at long times the behavior of both $\langle x^{2}\rangle $ and $%
\langle y^{2}\rangle $ is effectively "diffusive". This does not imply, of
course, that the processes $x(t)$ and $y(t)$ are standard Brownian motions
but rather signifies that such a behavior results from an interplay between
an ongoing input of energy (leading to a super-diffusive motion, Eq. %
\eqref{e}) and the restoring force due to the confining potential, which partially counter-balance each
other. Indeed, one notices that because of this  trade-off the prefactors in
the leading terms in Eqs. \eqref{x} and \eqref{y} are proportional to the
sum of the temperatures, and are inversely proportional to $%
\lambda $. In Sec. \ref{sec:twotime} we will discuss the ageing behavior of
the process $x(t)$ as embodied in its two-time covariance $\langle
x(t)x(t_{1})\rangle $, which manifests significant departures from the
standard Brownian motion.

Lastly, we find  the mixed moment  $\langle xy\rangle
$ of the components of instantaneous position. Using our Eqs. \eqref{a} and \eqref{b}, we find that for any $t>0$
\begin{equation}
\left\langle xy\right\rangle =-\frac{u\,\left( T_{1}+T_{2}\right) }{2\lambda
(1-u^{2})}\,t-\frac{\left( T_{1}+T_{2}\right) }{8}\left( \dfrac{\sin \left(
2\Omega _{+}t\right) }{\Omega _{+}^{3}}-\dfrac{\sin \left( 2\Omega
_{-}t\right) }{\Omega _{-}^{3}}\right) \,,
\label{xy}
\end{equation}%
We observe that the covariance depends only on the sum of the
temperatures, unlike the moments $\left\langle x^{2}\right\rangle$ and
$\left\langle y^{2}\right\rangle$ in Eqs. \eqref{x} and \eqref{y}. The leading large-%
$t$ term of Eq. \eqref{xy} grows linearly with time and its sign is opposed to
that of the coupling parameter $u$ and the sub-leading term oscillates irregularly.

\subsection{Second moments of velocities and velocity-position correlations.}

The second moments of the velocities $\dot{x}(t)$ and $\dot{y}(t)$ can be
straightforwardly evaluated from our Eqs. \eqref{c} and \eqref{d}. Skipping
the details of the intermediate calculations, we get
\begin{equation}
\begin{split}  \label{va}
\left \langle \dot{x}^2 \right \rangle &= \frac{\left(T_1+T_2\right)}{2 } \,
t + \frac{\left(T_1+T_2\right)}{8 } \left(\frac{\sin\left(2 \Omega_+ t
\right)}{\Omega_+} + \frac{\sin\left(2 \Omega_- t\right)}{\Omega_-}\right) \\
&+ \frac{\left(T_1-T_2\right)}{2 u \lambda} \Bigg[\Omega_+
\cos\left(\Omega_- t\right) \sin\left( \Omega_+ t\right) - \Omega_-
\cos\left(\Omega_+ t\right) \sin\left( \Omega_- t\right)\Bigg] \,,
\end{split}%
\end{equation}
and
\begin{equation}
\begin{split}  \label{vb}
\left \langle \dot{y}^2 \right \rangle &= \frac{\left(T_1+T_2\right)}{2} \,
t + \frac{\left(T_1+T_2\right)}{8 } \left(\frac{\sin\left(2 \Omega_+ t
\right)}{\Omega_+} + \frac{\sin\left(2 \Omega_- t\right)}{\Omega_-}\right) \\
&- \frac{\left(T_1-T_2\right)}{2 u \lambda} \Bigg[\Omega_+
\cos\left(\Omega_- t\right) \sin\left( \Omega_+ t\right) - \Omega_-
\cos\left(\Omega_+ t\right) \sin\left( \Omega_- t\right)\Bigg] \,.
\end{split}%
\end{equation}
We see that in the long time limit fluctuations of the
velocities grow "diffusively", i.e., the second moments of velocities
increase in proportion to the first power of time. The transient terms,
important at the intermediate stages, exhibit irregular oscillations, because the frequencies $\Omega_{\pm}$ are incommensurate generically. Such
an ultimate diffusive growth is very similar (apart of a dimensional
prefactor) to the one which we have previously observed for the second
moments of $x(t)$ and $y(t)$ (see Eqs. \eqref{x} and \eqref{y}). There is,
however, an important difference in the behavior of the moments of the position and
those  of the velocities. Namely, consider the limit $\lambda \to 0$, in
which case the potential vanishes and $x(t)$ and $y(t)$ are standard
independent random-acceleration processes that display a strongly
super-diffusive behavior, (see Eq. \eqref{e}). Setting $\lambda = 0$ in Eqs. %
\eqref{va} and \eqref{vb}, we get
\begin{equation}
\begin{split}
\langle \dot{x}^2 \rangle &= 2 T_1 \, t \,, \quad \langle \dot{y}^2 \rangle
= 2 T_2 \, t \,, \quad \lambda = 0 \,.
\end{split}%
\end{equation}
Next, for $\lambda > 0$ but $u = 0$, when the components $x(t)$ and $y(t)$
are two decoupled random-acceleration processes evolving each in a quadratic
potential, our Eqs. \eqref{va} and \eqref{vb} yield
\begin{equation}
\begin{split}
\langle \dot{x}^2 \rangle &= T_1 \, t + \frac{T_1}{2 \lambda^{1/2} }
\sin\left(2 \sqrt{\lambda} t\right) \,, \\
\langle \dot{y}^2 \rangle &= T_2 \, t + \frac{T_2}{2 \lambda^{1/2} }
\sin\left(2 \sqrt{\lambda} t\right) \,, \quad u = 0, \,\,\lambda > 0.
\end{split}%
\end{equation}
We  conclude that the diffusive growth of fluctuations of
velocities is a universal feature independent of the fact whether the potential
in Eq. \eqref{3} is present or not, while the fluctuations of positions
themselves behave very differently depending whether potential is present or
not.

Further on, we find that the covariance $\langle \dot{x} \dot{y} \rangle$
of the components of velocities is given by
\begin{equation}
\begin{split}  \label{ka}
\left \langle \dot{x} \dot{y} \right \rangle &= \frac{(T_1+T_2)}{8 \sqrt{%
\lambda(1 - u^2)}} \left( \frac{\sin\left(2 \Omega_+ t\right)}{\Omega_+} -
\frac{\sin\left(2 \Omega_- t\right)}{\Omega_-} \right) \,,
\end{split}%
\end{equation}
hence, do not grow with time (unlike the correlations of positions, Eq. \eqref{xy}%
), and oscillate around zero. The right-hand-side of Eq. \eqref{ka}
vanishes, as it should, when $\lambda \to 0$ or $u \to 0$.

Consider next the covariance of positions and velocities. One
readily finds
\begin{equation}
\begin{split}
\left \langle x \dot{x} \right \rangle &= \frac{1}{2}\frac{d\left \langle
x^2 \right \rangle}{dt} = \frac{\left(T_1 + T_2\right)}{ 4 } \, \left(\frac{%
\sin^2\left( \Omega_+ t\right)}{\Omega_+^2} + \frac{\sin^2\left( \Omega_-
t\right)}{\Omega_-^2} \right) \\
&- \frac{\left(T_1 - T_2\right)}{2 \lambda } \sin\left(\Omega_+ \,
t\right)\sin\left(\Omega_- t \right) \,,
\end{split}
\label{xxd}
\end{equation}
and
\begin{equation}
\begin{split}
\left \langle y \dot{y} \right \rangle &=\frac{1}{2}\frac{d\left \langle y^2
\right \rangle}{dt} = \frac{\left(T_1 + T_2\right)}{ 4 } \, \left(\frac{%
\sin^2\left( \Omega_+ t\right)}{\Omega_+^2} + \frac{\sin^2\left( \Omega_-
t\right)}{\Omega_-^2} \right) \\
&+ \frac{\left(T_1 - T_2\right)}{2 \lambda } \sin\left(\Omega_+
t\right)\sin\left(\Omega_- t\right) \,.
\end{split}
\label{yyd}
\end{equation}
Thus, the position-velocity covariances consist of two terms:
the first one, which is proportional to the sum of two temperatures, is
oscillating with time but is strictly positive and yields the positive
contribution to the covariance. The second term, which is
proportional to the difference of two temperatures and thus appears in
out-of-equilibrium situations only, is also oscillating and is changing its
sign with time. It is also easy to check that the position-velocity covariance is
always positive.

Lastly, we determine the position-velocity correlations of the form $%
\left\langle x\dot{y}\right\rangle $ and $\left\langle y\dot{x}\right\rangle
$. We find
\begin{equation}
\begin{split}
\left\langle x\dot{y}\right\rangle & =-\frac{u(T_{1}+T_{2})}{4\lambda
(1-u^{2})}+\frac{(T_{1}+T_{2})}{8}\left( \dfrac{\cos \left( 2\Omega
_{-}t\right) }{\Omega _{-}^{2}}-\dfrac{\cos \left( 2\Omega _{+}t\right) }{%
\Omega _{+}^{2}}\right) \\
& +\frac{(T_{2}-T_{1})}{2\lambda u}\left[ 1-\cos \left( \Omega _{-}t\right)
\cos \left( \Omega _{+}t\right) -\dfrac{\sin \left( \Omega _{-}t\right) \sin
\left( \Omega _{+}t\right) }{\sqrt{1-u^{2}}}\right] \,,
\end{split}
\label{k9}
\end{equation}%
and
\begin{equation}
\begin{split}
\left\langle y\dot{x}\right\rangle & =-\frac{u(T_{1}+T_{2})}{4\lambda
(1-u^{2})}+\frac{(T_{1}+T_{2})}{8}\left( \dfrac{\cos \left( 2\Omega
_{-}t\right) }{\Omega _{-}^{2}}-\dfrac{\cos \left( 2\Omega _{+}t\right) }{%
\Omega _{+}^{2}}\right) \\
& -\frac{(T_{2}-T_{1})}{2\lambda u}\left[ 1-\cos \left( \Omega _{-}t\right)
\cos \left( \Omega _{+}t\right) -\dfrac{\sin \left( \Omega _{-}t\right) \sin
\left( \Omega _{+}t\right) }{\sqrt{1-u^{2}}}\right] \,.
\end{split}
\label{k10}
\end{equation}%
Equations \eqref{k9} and \eqref{k10} show that the correlation function
between the either position and the velocity of the other component is a
bounded oscillating function of time which attains both negative and
positive values at different time moments. These moments vanish, as they
should, when either $\lambda \rightarrow 0$ or when $u\rightarrow 0$.

Together with the expressions for other the moments presented above they
determine completely the joint probability density $P_{t}$ and the
characteristic function $\Phi _{t}$ in Eqs. \eqref{dista} and \eqref{phit}.

\section{Two-time correlations}

\label{sec:twotime}

\begin{figure}
        \includegraphics[scale=0.44]{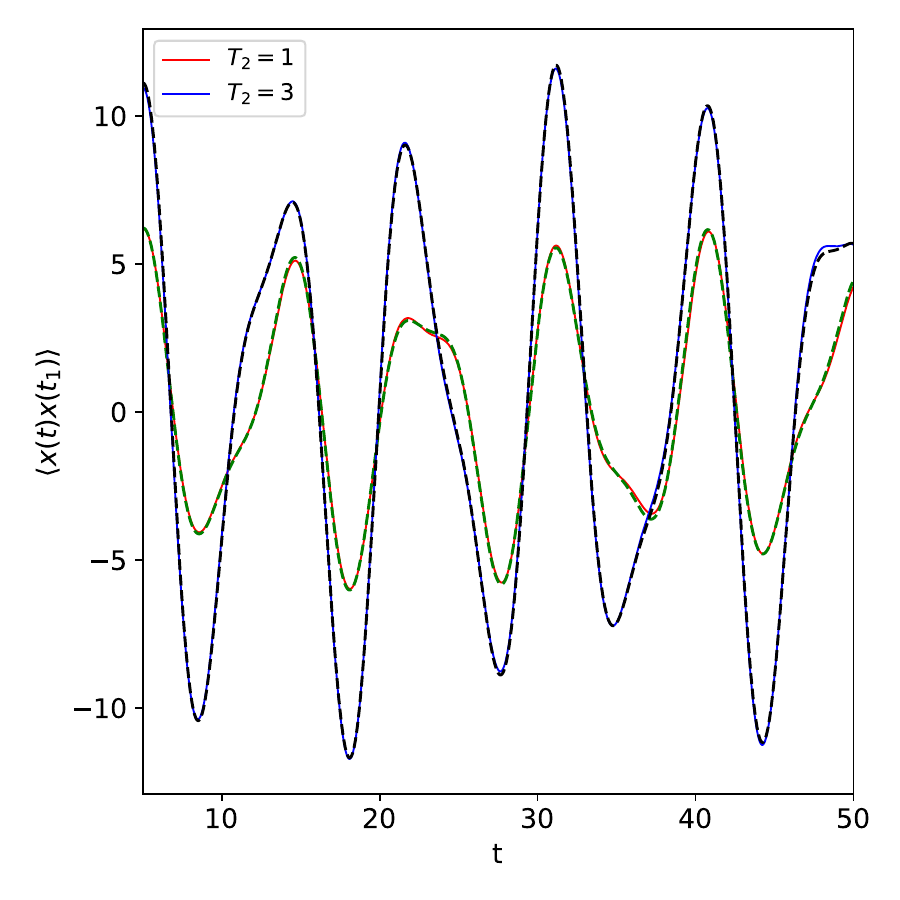}
        \includegraphics[scale=0.44]{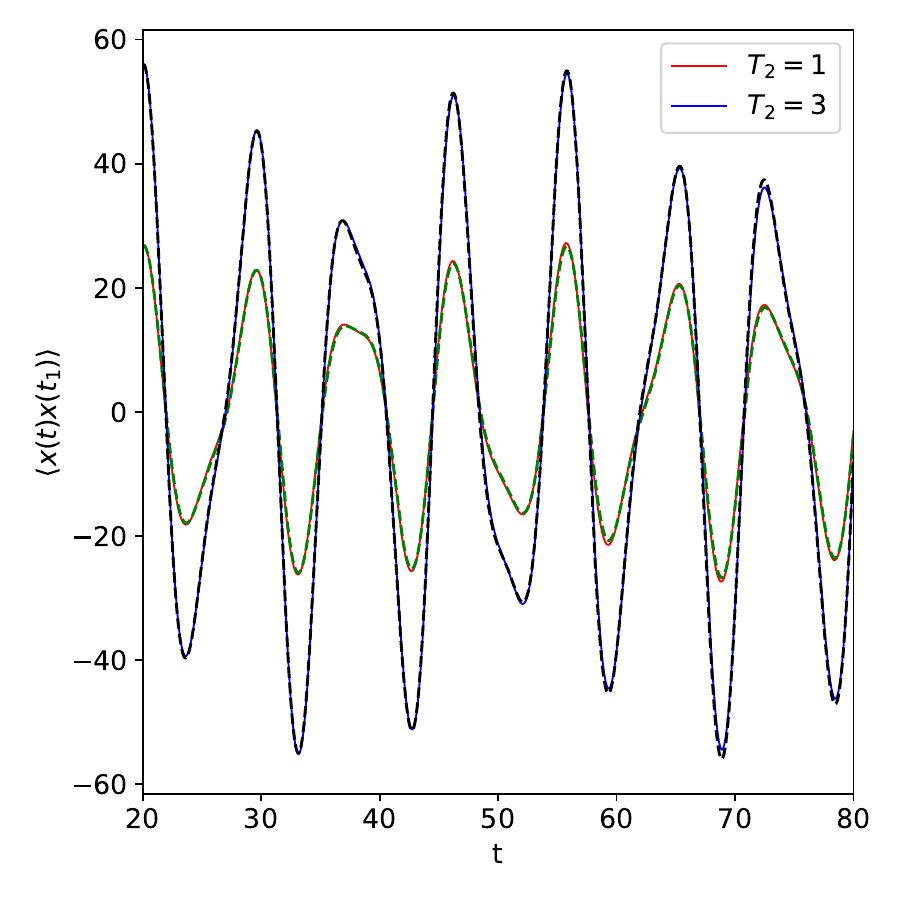}\\
        \caption{The covariance function $\langle x(t)x(t_1) \rangle$ versus time $t$ (for $t \geq t_1$) for two values of $t_1$ and temperatures $T_1=1$  and $T_2=1,3$. The coupling parameter $u=0.5$. Left panel:  $t_1=5$. Right panel: $t_1=20$. Dashed curves correspond to the exact expression, Eq. \eqref{eq:correlxx}.}\label{fig:correlxx}
\end{figure}

To get some additional insight into the time evolution of the processes
under study, say of $x(t)$, consider the two-time covariance $\langle
x(t)x(t_{1})\rangle $. Recall that the analogous covariance of standard Brownian
motion is $\langle x(t)x(t_{1})\rangle
=2D\min (t_{1},t)$ (e.g., $\langle x(t)x(t_{1})\rangle =2Dt_{1}$ for $%
t_{1}\leq t$) with $D$ being the diffusion coefficient. Using Eqs. \eqref{a}
and performing some straightforward calculations, we find that in our case
the covariance function for $t\geq t_{1}$ is
\begin{equation}
\begin{split}
\langle x(t)x(t_{1})\rangle & =\frac{\left( T_{1}+T_{2}\right) }{4}%
\,t_{1}\left( \frac{\cos \left( \Omega _{+}(t-t_{1})\right) }{\Omega _{+}^{2}%
}+\frac{\cos \left( \Omega _{-}(t-t_{1})\right) }{\Omega _{-}^{2}}\right)  \\
& +\frac{\left( T_{1}+T_{2}\right) }{4}\left( \frac{\sin \left( \Omega
_{+}t_{1}\right) \cos \left( \Omega _{+}t\right) }{\Omega _{+}^{3}}+\frac{%
\sin \left( \Omega _{-}t_{1}\right) \cos \left( \Omega _{-}t\right) }{\Omega
_{-}^{3}}\right)  \\
& -\frac{\left( T_{1}-T_{2}\right) }{4u\lambda }\left[ \Big(\cos \left(
\Omega _{+}\,t_{1}\right) -\cos \left( \Omega _{-}\,t_{1}\right) \Big)\left(
\frac{\sin \left( \Omega _{+}t\right) }{\Omega _{+}}+\frac{\sin \left(
\Omega _{-}t\right) }{\Omega _{-}}\right) \right.  \\
& \left. +\Big(\cos \left( \Omega _{+}\,t\right) +\cos \left( \Omega
_{-}\,t\right) \Big)\left( \frac{\sin \left( \Omega _{-}t_{1}\right) }{%
\Omega _{-}}-\frac{\sin \left( \Omega _{+}t\right) }{\Omega _{+}}\right) %
\right]
\end{split}%
\label{eq:correlxx}
\end{equation}%
Viewing $t_1$ as a parameter and $t$ as a variable, we observe that for sufficiently large $t_{1}$ (but $t\geq t_{1}$), the dominant
contribution comes from the terms in the first line of the r.h.s. of Eq. %
\eqref{eq:correlxx}, while the terms in the second and the third lines show
a purely oscillatory behavior and are bounded function of both $t$ and $t_{1}
$. The amplitude of the dominant terms is proportional to the sum of the
temperatures and to $t_{1}$, (similarly to the standard Brownian motion). In
contrast to the Brownian motion, the amplitude is multiplied by a function
of $t-t_{1}$, which exhibits irregular oscillations. The covariance function
is depicted in Fig.~\ref{fig:correlxx} as a function of $t$ for two values
of $t_{1}$ and two values of the temperatures.

\section{Position-velocity probability densities}

\label{sec:Pt}

From the joint probability density $P_{t}$ in Eq. \eqref{dista} or the characteristic
function $\Phi_{t}$, Eq. \eqref{phit}, together with the exact
expressions for the moments and the mixed moments of position and velocity
derived in Sec. \ref{sec:3}, we can readily calculate
the marginal position-velocity distributions $P_{t}(x,\dot{x})$ and $P_{t}(y,\dot{y})$.
Performing the corresponding Gaussian integrations, we find
\begin{equation}
P_{t}(x,\dot{x})=\frac{1}{2\pi \sqrt{\langle x^{2}\rangle \langle \dot{x}%
^{2}\rangle -\langle x\dot{x}\rangle ^{2}}}\exp \left( -\frac{\langle \dot{x}%
^{2}\rangle x^{2}-2\langle x\dot{x}\rangle x\dot{x}+\langle x^{2}\rangle
\dot{x}^{2}}{2\left( \langle x^{2}\rangle \langle \dot{x}^{2}\rangle
-\langle x\dot{x}\rangle ^{2}\right) }\right) \,,  \label{distb}
\end{equation}%
and
\begin{equation}
P_{t}(y,\dot{y})=\frac{1}{2\pi \sqrt{\langle y^{2}\rangle \langle \dot{y}%
^{2}\rangle -\langle y\dot{y}\rangle ^{2}}}\exp \left( -\frac{\langle \dot{y}%
^{2}\rangle y^{2}-2\langle y\dot{y}\rangle y\dot{y}+\langle y^{2}\rangle
\dot{y}^{2}}{2\left( \langle y^{2}\rangle \langle \dot{y}^{2}\rangle
-\langle y\dot{y}\rangle ^{2}\right) }\right) \,.  \label{distc}
\end{equation}%
Equations \eqref{dista}, \eqref{distb} and \eqref{distc} are exponentials of
certain quadratic forms of the variables, i.e., are Gaussian as they should be,
but the  coefficients of the forms are rather complicated functions of the moments. This
makes difficult to write them explicitly and we confine ourselves only to their
large-time forms
\begin{equation}
\begin{split}
& P_{t\rightarrow \infty }(x,\dot{x})\simeq \frac{\sqrt{\lambda (1-u^{2})}}{%
\pi (T_{1}+T_{2})t}\exp \Bigg(-\frac{\lambda (1-u^{2})x^{2}}{(T_{1}+T_{2})t}-%
\frac{\dot{x}^{2}}{(T_{1}+T_{2})t}+ \\
& +\Bigg[\lambda \left( \frac{\sin ^{2}(\Omega _{+}t)}{\Omega _{+}^{2}}+%
\frac{\sin ^{2}(\Omega _{-}t)}{\Omega _{-}^{2}}\right) -\frac{2(T_{1}-T_{2})%
}{(T_{1}+T_{2})}\sin \left( \Omega _{+}t\right) \sin \left( \Omega
_{-}t\right) \Bigg]\frac{x\,\dot{x}}{(T_{1}+T_{2})t^{2}}\Bigg)\,,
\end{split}%
\label{distd}
\end{equation}%
and
\begin{equation}
\begin{split}
& P_{t\rightarrow \infty }(y,\dot{y})\simeq \frac{\sqrt{\lambda (1-u^{2})}}{%
\pi (T_{1}+T_{2})t}\exp \Bigg(-\frac{\lambda (1-u^{2})y^{2}}{(T_{1}+T_{2})t}-%
\frac{\dot{y}^{2}}{(T_{1}+T_{2})t}+ \\
& +\Bigg[\lambda \left( \frac{\sin ^{2}(\Omega _{ +}t)}{\Omega _{+}^{2}}+%
\frac{\sin ^{2}(\Omega _{-}t)}{\Omega _{-}^{2}}\right) +\frac{2(T_{1}-T_{2})%
}{(T_{1}+T_{2})}\sin \left( \Omega _{+}t\right) \sin \left( \Omega
_{-}t\right) \Bigg]\frac{y\,\dot{y}}
{(T_{1}+T_{2})t^{2}}\Bigg)\,.
\end{split}%
\label{diste}
\end{equation}%
We observe that the quadratic terms in the exponents in Eqs. \eqref{distd} and \eqref{diste} %
 contain the first
inverse power of time, as can be expected from the "diffusive" growth of the
second moments of positions and  the velocities (see Eqs. \eqref{x} and \eqref{y}, %
 and also Eqs. \eqref{va} and \eqref{vb}), while the amplitudes of the cross-terms
decay at a faster rate proportional to $1/t^{2}$, showing  that the
velocities effectively decouple from the positions in the large-$t$ limit.
The amplitudes of the cross-terms also contain oscillatory functions that change (aperiodically) their sign with time. We
note that this behavior also persists for the total joint probability density $%
P_{t}(x,y,\dot{x},\dot{y})$ in Eq. \eqref{dista}. However, the explicit
expression for the corresponding exponent  is quite cumbersome even in the
limit $t\rightarrow \infty $ and we do not present it here. We only
mention that, in contrast to the Brownian gyrator model, the limiting forms
of $P_{t}(x,y,\dot{x},\dot{y})$  and its  marginals (Eqs. \eqref{distd} and \eqref{diste}%
)   do not exist
for the two-dimensional random-acceleration process at $t= \infty$.

\section{Gyration characteristics: angular momentum and angular velocity}

\label{sec:gyration}

Similarly to the standard analyses of the BGM, we will study now the
behavior of the angular momentum $L(t)$, Eq. \eqref{meanL}, and of the angular
velocity $W(t)$, Eq. \eqref{meanomega}. 
Recall that the former, which is also
called the rotational moment, determines the torque exerted on the particle
being at some given instantaneous position, while the latter represents the
rate at which the particle at position $\mathbf{r}(t)$ rotates around a fixed origin.
Most of available analyses concentrated on the \textit{mean} values of these
characteristic properties in the limit $t\rightarrow \infty $; both attain
finite non-zero values when $T_{1}\neq T_{2}$ and equal zero in equilibrium
conditions $T_{1}=T_{2\text{ }}$(see, e.g. \cite%
{rei,Dotsenko2013,Mancois2018,Bae2021}). On this basis, it was often
concluded that the Brownian gyrator is a kind of a "nano-machine" that
steadily gyrates around the origin. A recent work \cite{viot2023destructive}
has questioned this conclusion by looking on the behavior beyond the mean
values. In particular, the probability densities of $L$ and $W$ have been
calculated and it was shown that the latter exist only for a
time-discretized (with time-step $\delta t$) version of the model. In
particular, it was also shown that for finite $\delta t$ the probability
density $P_{t}(L)$ of the angular momentum is always sharply peaked at $L=0$%
, and has the exponential tails with different slopes for $T_{1}\neq T_{2}$.
This implies that moments of angular momentum of arbitrary order exist, but
their values are supported by the tails of the distribution and
consequently, do not represent the typical behavior $L$. Moreover, it was
demonstrated that the variance of $L$ is always much larger than the squared
first moment, i.e., the "noise" is always greater than the "signal", and
diverges in the limit $\delta t\rightarrow 0$. More strikingly, the
probability density $P_{t}(W)$ of the angular velocity has
algebraic large-$W$ tails of the form $P_{t}(W)\simeq 1/|W|^{3}$ such that,
in fact, the first moment is the \textit{only} existing moment. This
signifies that the spread of the values of angular velocities is infinitely
large, although for a large ensemble of "gyrators" there exists some
non-zero averaged value of the velocity in out-of-equilibrium conditions. In
the limit $\delta t\rightarrow 0$, both probability densities
converge to uniform distributions with a vanishing amplitude and diverging
variance.

We will analyze below the behavior of the moments of the angular
momentum and velocity, as well as their probability densities for a
random-acceleration process under study along exactly the same lines as it
was for the BGM.

\subsection{Angular momentum}

\begin{figure}[th]
\includegraphics[width=0.5\textwidth,clip=]{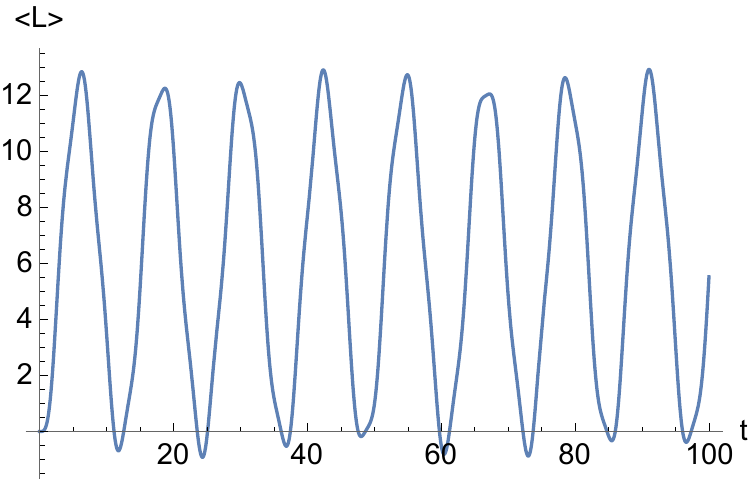} %
\includegraphics[width=0.5\textwidth,clip=]{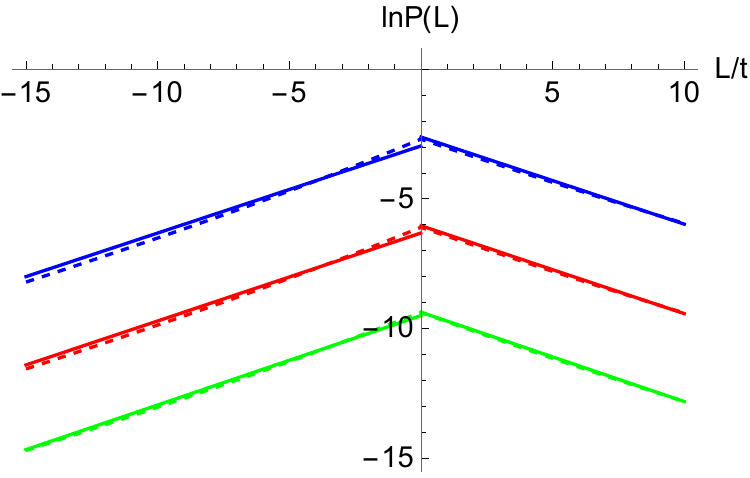}
\caption{Angular momentum. Left panel: Mean angular momentum $\langle
L\rangle $ versus time for $T_{1}=1$, $T_{2}=4$, $u=1/2$ and $\protect%
\lambda =1$. Solid curve depicts the exact result in Eq. \eqref{LLmean}.
Right panel: Logarithm of the probability density  $P_{t}(L)$ as function of scaled variable $L/t$ for $T_{1}=1$, $T_{2}=4$, $u=1/2$ and
$\protect\lambda =1$. Dashed curves correspond to a numerical evaluation of
the integral in Eq. \eqref{PL} for $t=20$ (blue), $t=30$ (red) and $t=40$
(green). Solid curves with the same color code correspond to the asymptotic
large-$L$ form in Eq. \eqref{PLas}. Note that since all the curves appear
too close to each other in the plot, for notational convenience we shifted
upwards both blue curves by $+2$, both red curves are shifted downwards by $%
-1$ units, and the green curves - by $-4$. }
\label{fig:L}
\end{figure}

 According to Eq. \eqref{meanL}, the mean value of the angular momentum is
 given by (recall that we have set $m=1$)
\begin{equation}\label{Lmean}
\left\langle L\right\rangle =\langle x\dot{y}\rangle -\langle y\dot{x}%
\rangle \,,
\end{equation}%
where the terms in the r.h.s. are given by Eqs. \eqref{k9} and %
\eqref{k10}. Performing some simple calculations, we find that the
first moment of the angular momentum obeys
for any $t>0$
\begin{equation}\label{LLmean}
\left\langle L\right\rangle =\frac{(T_{2}-T_{1})}{\lambda u}\Big(1-\cos
(\Omega _{+}t)\cos (\Omega _{-}t)-\frac{\sin (\Omega _{+}t)\sin (\Omega
_{-}t)}{\sqrt{1-u^{2}}}\Big)\,.%
\end{equation}%
 Therefore, the first moment is not identically equal to  zero if $T_{1}\neq T_{2}$. Moreover, it is easy to show that the first moment is an odd function of $u$, and vanishes when $u=0$. This latter case corresponding to two independent random acceleration processes, such that no angular momentum is expected. However, in
contrast to the BGM, $\left\langle L\right\rangle $ does not approach a
constant value as $t\rightarrow \infty $, but rather exhibits irregular
oscillations with time and can be positive or negative at different time
moments. This means that the torque exerted on the particle changes the
direction at different time moments and the fraction of time that $%
\left\langle L\right\rangle $ has positive values is controlled by the sign
of the temperature difference $T_{2}-T_{1}$. For $T_{2}-T_{1}>0$, the mean
value of the angular moment is predominantly positive, as one can observe
from the plot presented on the left panel in Fig. \ref{fig:L}.

Consider next the fluctuations of $L$ around its mean value. To this end, it is
convenient to use the generalized Wick theorem, according to which if $%
X=\{X_{j}\}_{j=1}^{m}$ is the collection of Gaussian random variables with
zero mean and $f$ is a function of $X$, then we have for $j=1,2,\ldots ,m$%
\begin{equation*}
\langle X_{j}f(X)\rangle =\sum_{k=1}^{m}\langle X_{j}X_{k})\rangle
\left\langle X_{k}\frac{\partial f}{\partial X_{k}}\right\rangle .
\end{equation*}%
The theorem follows readily from Eqs. \eqref{dista} or \eqref{phit}). This implies, in
view of Eq. \eqref{meanL}, that
\begin{equation}
\begin{split}
\left\langle L^{2}\right\rangle & =2\left\langle x\dot{y}\right\rangle
^{2}+2\left\langle \dot{x}y\right\rangle ^{2}+\left\langle
x^{2}\right\rangle \left\langle \dot{y}^{2}\right\rangle +\left\langle \dot{x%
}^{2}\right\rangle \left\langle y^{2}\right\rangle  \\
& -2\Big(\left\langle xy\right\rangle \left\langle \dot{x}\dot{y}%
\right\rangle +\left\langle x\dot{y}\right\rangle \left\langle \dot{x}%
y\right\rangle +\left\langle x\dot{x}\right\rangle \left\langle y\dot{y}%
\right\rangle \Big)\,.
\end{split}%
\end{equation}%
This and Eq. \eqref{Lmean}  yield the following expression for the variance of the angular momentum
\begin{equation}
\begin{split}
\mathrm{Var}\{L\}=\left\langle L^{2}\right\rangle -\left\langle
L\right\rangle ^{2}& =\left\langle x\dot{y}\right\rangle ^{2}+\left\langle
\dot{x}y\right\rangle ^{2}+\left\langle x^{2}\right\rangle \left\langle \dot{%
y}^{2}\right\rangle +\left\langle \dot{x}^{2}\right\rangle \left\langle
y^{2}\right\rangle  \\
& -2\left\langle x\dot{x}\right\rangle \left\langle y\dot{y}\right\rangle
-2\left\langle xy\right\rangle \left\langle \dot{x}\dot{y}\right\rangle \,,
\end{split}%
\end{equation}%
 which is valid for any $t$. Then, by using the results of Sec. \ref{sec:3}, we determine the leading term of
the large-$t$ asymptotic form
\begin{equation}
\mathrm{Var}\{L\} \simeq \frac{(T_{1}+T_{2})^{2}}{2\lambda
(1-u^{2})}t^{2}\,.
\end{equation}%
It follows from Eq. \eqref{LLmean}  that $|\langle L\rangle |\leq
3|T_{2}-T_{1}|(\lambda u(1-u^{2})^{1/2})^{-1}$, and hence, we find that the
relative mean square deviation of $L$  (i. e., the coefficient of variation of the corresponding probability density, see, e.g.,  \cite{mejia}) obeys the inequality:
\begin{equation}
\frac{(\mathrm{Var}\left\{ L\right\} )^{1/2}}{|\langle L\rangle |}\geq Ct,\;C=%
\frac{(T_{1}+T_{2})u\sqrt{\lambda }}{|T_{1}-T_{2}|3\sqrt{2}}.  \label{msd}
\end{equation}%
Thus, the relative fluctuations of $L$ grow at least linearly in $t$. Hence, the first
moment of the angular momentum does not have much of a physical significance and
only indicates some trend in the statistical ensemble.

We conclude that in contrast to many-body physics, where macroscopic
observables do not fluctuate and are therefore completely characterized by
their mean values, the angular momentum, (as well as the angular velocity, as we will demonstrate below), in our
model do not have this property (known as representativeness of averages
in statistical physics and self-averaging in disordered media physics).
Therefore, in order to obtain complete information about such strongly
fluctuating observables, it is necessary to have their complete probability
distribution.

This is why we will turn now to the analysis of the probability density of the
angular momentum $L$, Eq. \eqref{meanL}. We will begin with the characteristic function of $L$
\begin{equation}
\Phi _{L}(\nu)=\left\langle \exp \left( i\nu \left( x\dot{y}-y\dot{x%
}\right) \right) \right\rangle .
\label{La0}
\end{equation}%
It is shown Appendix \ref{secA1} that
\begin{equation}
\Phi _{L}(\nu )=\frac{1}{\nu^{2}\sqrt{\mathrm{det}(M_{\nu })}}
\label{La1}
\end{equation}%
where $M_{\nu}$ is a $4\times 4$ matrix
\begin{equation}\label{eq:Mat}
M_{\nu}=M(t)+\frac{1}{\nu}\left(
\begin{array}{cc}
0 & -\sigma _{y} \\
\sigma _{y} & 0%
\end{array}%
\right) ,\;\sigma _{y}=\left(
\begin{array}{cc}
0 & -i \\
i & 0%
\end{array}%
\right)
\end{equation}
and $M(t)$ is given in Eq. \eqref{eq:MatM}.

Therefore, the characteristic function $\Phi _{L}(\nu)$ is the
inverse of a square root of a quartic polynomial in $\nu$, whose
coefficients can be expressed via the moments of $L$. In particular, it can be shown that the
coefficient in front of $\nu$ is $-2i\langle L\rangle $, the
coefficient in front of $\nu^{2}$ is $(\langle L^{2}\rangle -3\langle
L\rangle ^{2})$, etc.

Given $\Phi _{L}(\nu)$, the probability density $%
P_{t}(L)$ of $L$ is:
\begin{equation}
P_{t}(L)=\frac{1}{2\pi }\int_{-\infty }^{\infty }\frac{d\nu}{\nu
^{2}\sqrt{\mathrm{det}(M_{\nu})}}\exp (-i\nu L)\,.  \label{PL}
\end{equation}%
Despite a relatively simple form of the integrand, the above integral  cannot be performed exactly. We hence resort to a
numerical evaluation of this integral and its asymptotic analysis in the
limit $|L|\rightarrow \infty $.

Figure \ref{fig:L} presents the results of a numerical evaluation of $%
P_{t}(L)$ in Eq. \eqref{PL}, together with its asymptotic forms (see Eq. %
\eqref{PLas} below). For convenience, we plot the logarithm of $%
P_{t}(L)$ versus a scaled variable $L/t$. Numerically evaluated $%
P_{t}(L)$ is depicted by dashed curves for three values of $t$ : $t=20$
(blue), $t=30$ (red) and $t=40$ (green). We observe that $P_{t}(L)$ has a
cusp at $L=0$ where it attains the maximal value, similarly to the case of
the BGM. Consequently, the first moment of $L$ does not correspond to the
typical behavior, when $T_{1}\neq T_{2}$, and is therefore supported by the
whole tails of $P_{t}(L)$ (cf. Eq. \eqref{msd}). For larger $L$, both right and
left tails of the probability density seem to be exponential
functions of $L/t$. In order to verify if this is indeed the case, we turn
to Eq. \eqref{PL} and change the integration variable $\nu  \rightarrow
\nu /t$. Then, we find that the terms with the odd powers of $\nu $
in the denominator vanish as $t\rightarrow \infty $, and the term $\nu
^{4}$ is relevant only for the short-$L$ behavior. Discarding this latter
term, we get the following large-$L$ asymptotic formula
\begin{equation}
\begin{split}
P_{t}(L)& =\frac{\sqrt{\lambda (1-u^{2})}}{(T_{1}+T_{2})\,t}\exp \left( -%
\frac{2\sqrt{\lambda (1-u^{2})}}{(T_{1}+T_{2})}\frac{|L|}{t}\right)  \\
& \times \left( 1+\frac{2\sqrt{\lambda (1-u^{2})}}{(T_{1}+T_{2})t}\langle
L\rangle \,\mathrm{sign}(L)+O\left( \frac{1}{t^{2}}\right) \right) \,.
\end{split}%
\label{PLas}
\end{equation}%
Hence, the large-$L$ tails of the probability density of the angular
momentum are indeed exponential functions of $L$. Moreover, the left and
right tails have the same slope so that $P_{t}(L)$ for sufficiently large $L$
and $t$ is a symmetric function of $L$ with respect to $L=0$. This is quite
different from the behavior of probability density of $L$ observed in the
BGM, where the right and left tails have different slopes \cite%
{viot2023destructive}. The maximal value $P_{t}(L=0)$ decays as $t^{-1}$ and
the slope of the tails tends to zero as $t\rightarrow \infty $, which
explains why the variance of $L$ grows in proportion to $t^{2}$. The
asymptotic form of Eq. \eqref{PLas} is depicted in Fig. \eqref{fig:L} and we
observe that already for $t=40$ it becomes almost indistinguishable from the
numerical result.

\subsection{Angular velocity}

As in the above case of the angular momentum, it is convenient to use the
characteristic function, Eq. \eqref{phit}. In particular, the mean value of the
angular velocity (see Eq. \eqref{meanomega} with $m=1$) can be conveniently
represented as
\begin{equation}
\begin{split}
\left\langle W\right\rangle & =\frac{1}{4\pi }\int_{0}^{\infty }\frac{d\xi }{%
\xi }\int_{-\infty }^{\infty }\int_{-\infty }^{\infty }d\omega _{1}d\omega
_{2}\exp \left( -\frac{\omega _{1}^{2}+\omega _{2}^{2}}{4\xi }\right)  \\
& \times \left. \left\{ \left( -\frac{\partial ^{2}}{\partial \omega
_{1}\partial \omega _{4}}+\frac{\partial ^{2}}{\partial \omega _{2}\partial
\omega _{3}}\right) \Phi _{t}(\boldsymbol{\omega })\right\} \right\vert
_{\omega _{3}=\omega _{4}=0}\,,
\end{split}%
\end{equation}%
where the derivatives with respect to $\omega $-s give  the angular
momentum, while the integrations over $\omega _{1}$ and $\omega _{2}$, and
eventually, over $\xi $, produce the first inverse power of the moment of
inertia $x^{2}+y^{2}$. Performing straightforward calculations, we get for
any $t>0$ (cf. Eq. \eqref{LLmean})%
\begin{equation}
\begin{split}
\left\langle W\right\rangle & =-\frac{\left( \langle x^{2}\rangle +\langle
y^{2}\rangle -2\sqrt{\langle x^{2}\rangle \langle y^{2}\rangle -\langle
xy\rangle ^{2}}\right) }{\left( 4\langle xy\rangle ^{2}+\left( \langle
x^{2}\rangle -\langle y^{2}\rangle \right) ^{2}\right) \sqrt{\langle
x^{2}\rangle \langle y^{2}\rangle -\langle xy\rangle ^{2}}}\Big(\langle
xy\rangle \left( \langle y\dot{y}\rangle -\langle x\dot{x}\rangle \right)  \\
& +\langle y\dot{x}\rangle \left( \langle x^{2}\rangle +\sqrt{\langle
x^{2}\rangle \langle y^{2}\rangle -\langle xy\rangle ^{2}}\right) -\langle x%
\dot{y}\rangle \left( \langle y^{2}\rangle +\sqrt{\langle x^{2}\rangle
\langle y^{2}\rangle -\langle xy\rangle ^{2}}\right) \Big)\,,
\end{split}%
\label{omegamean}
\end{equation}%
The moments entering the above expression are given in Sec. \ref{sec:3}.

\begin{figure}[ht]
\begin{center}
\includegraphics[width=0.62\textwidth,clip=]{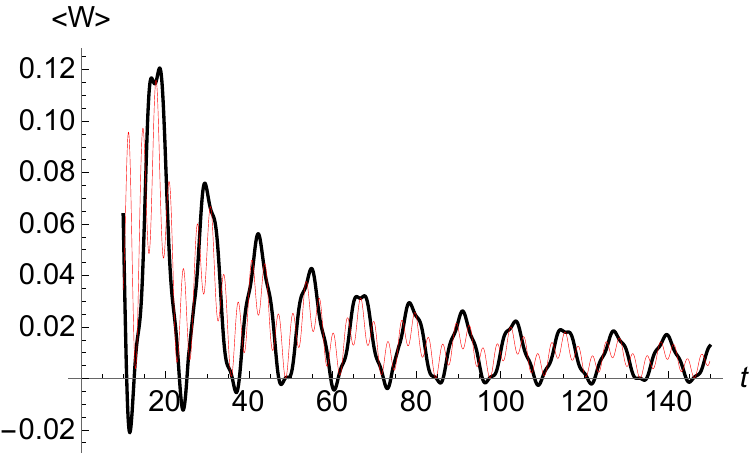}\\[0pt]
\includegraphics[width=0.62\textwidth,clip=]{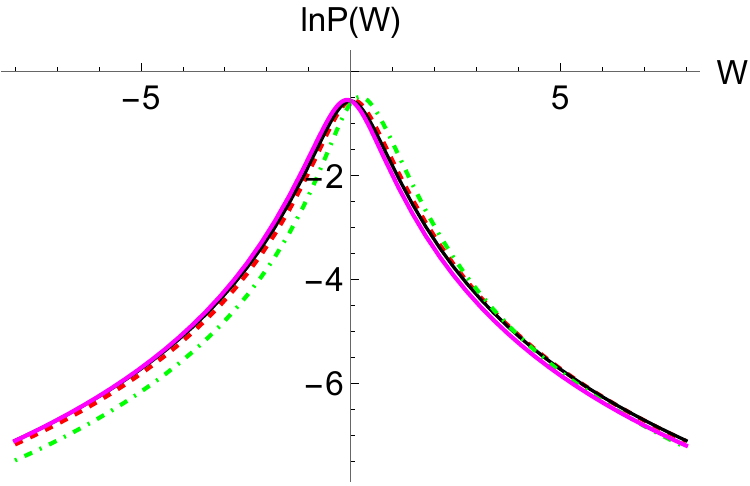}s
\end{center}
\caption{Angular velocity. Upper panel: Mean angular velocity $\langle W
\rangle$ versus time $t$ for $T_1=1$, $T_2 = 4$, $u = 1/2$ and $\protect%
\lambda = 1$. Solid (black) curve depicts the exact result in Eq.
\eqref{omegamean}, while the thin (red) curve presents the large-$t$
asymptotic form in Eq. \eqref{asomega}. Lower panel: Logarithm of the
probability density $P(W)$ as function of angular velocity for $%
T_1=1$, $T_2 = 4$, $u = 1/2$ and $\protect\lambda = 1$. Dashed green and red
curves are the exact results for $t = 5$ and $t =30$, respectively, obtained
by a numerical evaluation of expression \eqref{omegapdf}. Black solid line
corresponds to the limiting large-$t$ form in Eq. \eqref{zu}. Solid
(magenta) curve depicts the exact result in Eq. \eqref{omegapdf} for
two distinctly different temperatures $T_1 = 50$ and $T_2 = 1$. }
\label{fig:omega}
\end{figure}

In the large-$t$ limit, Eq. \eqref{omegamean} simplifies considerably to
give (cf. Eq. \eqref{LLmean})
\begin{equation}
\left\langle W\right\rangle \simeq \frac{(T_{2}-T_{1})\sqrt{1-u^{2}}}{%
(T_{1}+T_{2})ut}\Big(1-\cos \left( \Omega _{+}t\right) \cos \left( \Omega
_{-}t\right) \Big)\,.  \label{asomega}
\end{equation}%
Likewise the first moment  of $L$, Eq. \eqref{LLmean}, the mean value of $W$
is not identically  equal to zero only in "out-of-equilibrium" case where $%
T_{1}\neq T_{2}$; it is also  an odd function of $u$ that vanishes for $u=0$. As shown in Fig. \ref{fig:omega} (upper panel), $\langle
W\rangle $ is an oscillatory function of time whose envelope first rises to
some peak value and then decays as the first inverse power of time. The decay stems from the fact that the particle becomes delocalized and
moves away from the origin with time. In addition, similarly to $\langle
L\rangle$,  $\langle
W\rangle $ assumes both positive and negative values. Recall that in the standard BGM the mean
angular velocity approaches a constant value in the limit $t\rightarrow
\infty $ whose sign is defined by the temperature difference.

Consider now the second moment of the angular velocity. This moment can be
formally represented as
\begin{equation}
\begin{split}
\left\langle W^{2}\right\rangle & =\frac{1}{4\pi }\int_{0}^{\infty }d\xi
\int_{-\infty }^{\infty }\int_{-\infty }^{\infty }d\omega _{1}d\omega
_{2}\exp \left( -\frac{\omega _{1}^{2}+\omega _{2}^{2}}{4\xi }\right)  \\
& \times \left. \left\{ \left( \frac{\partial ^{4}}{\partial \omega
_{1}^{2}\partial \omega _{4}^{2}}+\frac{\partial ^{4}}{\partial \omega
_{2}^{2}\partial \omega _{3}^{2}}-2\frac{\partial ^{4}}{\partial \omega
_{1}\partial \omega _{2}\partial \omega _{3}\partial \omega _{4}}\right)
\Phi _{t}(\boldsymbol{\omega })\right\} \right\vert _{\omega _{3}=\omega
_{4}=0}\,,
\end{split}%
\end{equation}%
where the differential operator yields the second power of $L$
, while the integration over $\xi $ gives here the second
inverse power of the moment of inertia. Performing differentiations, and
then integrating over $\omega _{1}$ and $\omega _{2}$, we find some
complicated function of $\xi $. Analyzing the decay of this function in the
large-$\xi $ limit, we find that it vanishes as $1/\xi $. This signifies
that the integral over $\xi $ diverges logarithmically at infinity, hence,
\begin{equation}
\left\langle W^{2}\right\rangle =\infty \,,
\end{equation}%
which is precisely the behavior encountered previously in the BGM.
Consequently, the probability density of the angular velocity has \textit{heavy}
power-tails such that the mean angular velocity is the only existing moment.

We turn therefore to the analysis of the probability density of the angular
velocity. As in the previous subsection, we consider first the
characteristic function of the angular velocity and we find
\begin{equation}
\begin{split}
\Phi _{W}(z)& =\left\langle \exp \left( iz\frac{\left( x\dot{y}-y\dot{x}%
\right) }{x^{2}+y^{2}}\right) \right\rangle  \\
& =\frac{|z|}{2\pi \sqrt{d}}\int_{0}^{2\pi }d\theta \left( \frac{C_{\theta }%
}{A_{\theta }}\right) ^{1/2}K_{1}\left( \frac{\sqrt{A_{\theta }C_{\theta }}}{%
d}|z|\right) \,\exp \Big(-\frac{iB_{\theta }z}{2d}\Big)\,,
\end{split}
\label{omegax}
\end{equation}%
where $K_{1}(x)$ is the modified Bessel function,
\begin{equation}
d=\langle x^{2}\rangle \langle y^{2}\rangle -\langle xy\rangle ^{2}\,,
\end{equation}%
while the coefficients $A_{\theta }$, $B_{\theta }$ and $C_{\theta }$ are
functions of $\theta $ and $t$ but do not depend on $W.$ These functions (see
Eqs. \eqref{cts}), as well as the details of intermediate calculations are
presented in Appendix \ref{secA2}.

Inverting the Fourier transform, we find next that the probability density
function of the angular velocity valid for any $t$ admits the integral
representation
\begin{equation}
P_{t}(W)=\frac{2d^{3/2}}{\pi }\int_{0}^{2\pi }\frac{C_{\theta }\,d\theta }{%
\left( 4A_{\theta }C_{\theta }+\left( B_{\theta }+2dW\right) ^{2}\right)
^{3/2}}
\label{omegapdf}%
\end{equation}%
A simple analysis shows that the large-$W$ asymptotic form of $P_{t}$ is
\begin{equation}
P_{t}(W)\simeq \frac{\alpha }{|W|^{3}}\,,\;\alpha =\frac{1}{4\pi d^{3/2}}%
\int_{0}^{2\pi }C_{\theta } \, d\theta .  \label{Was}
\end{equation}%
i.e., $\alpha $ depends on time, both temperatures and the parameters
characterizing the potential (see
Eqs. \eqref{cts}). In Fig. \ref{fig:omega} we depict the results
of a numerical evaluation of the integral in Eq. \eqref{omegapdf} together
with the asymptotic form in Eq. \eqref{Was} (see below).

Lastly, we concentrate on the behavior of $P_{t}(W)$ in Eq. \eqref{omegapdf}
in the limit $t\rightarrow \infty $ and $W$ bounded away from zero, which
appears to be quite non-trivial. To this end, we observe that in the limit $%
t\rightarrow \infty $ the function $B_{\theta }$ (see Eqs. \eqref{cts})
grows linearly with time, i.e., $B_{\theta }\simeq t\,\Gamma _{\theta }$,
where $\Gamma _{\theta }$ is a rather complicated function of $\theta $, $u$%
, $\lambda $ and both temperatures, but is independent of time. On the other
hand, the asymptotic behavior of the functions
\begin{equation}
\begin{split}
& A_{\theta }\simeq \frac{(T_{1}+T_{2})(1+u\sin (2\theta ))}{2\lambda
(1-u^{2})}\,t\,, \\
& C_{\theta }\simeq \frac{(T_{1}+T_{2})^{3}}{8\lambda ^{2}(1-u^{2})}%
\,t^{3}\,, \\
& d\simeq \frac{(T_{1}+T_{2})^{2}}{4\lambda ^{2}(1-u^{2})}t^{2}\,,
\end{split}%
\end{equation}%
is given by explicit and fairly compact expressions. Inserting these
expressions into Eq. \eqref{omegapdf}, we find that $P_{t}(W)$ attains the
following limiting ($t=\infty $) form, valid for any $W\neq 0$ :
\begin{equation}
\begin{split}
P_{\infty }(W)\simeq\frac{\lambda \sqrt{1-u^{2}}}{8\pi \left( \lambda
+W^{2}\right) ^{3/2}}\int_{0}^{\pi }\frac{d\theta }{\left( 1+\kappa \sin
(\theta )\right) ^{3/2}}\,,\quad \kappa =\frac{\lambda u}{\lambda +W^{2}}\,.
\end{split}%
\label{zu}
\end{equation}%
The integral in Eq. \eqref{zu} can be expressed through the elliptic
integrals. But its large-$W$ asymptotic form follows directly from Eq. \eqref{zu}%
 , (e.g., by expanding the integrand into the Taylor series in powers of $\kappa$):
\begin{equation}
P(W)\simeq\frac{\lambda \sqrt{1-u^{2}}}{8|W|^{3}}\,,W\rightarrow \pm \infty \,,
\label{zut}
\end{equation}%
(cf. Eq. \eqref{Was}). Curiously enough, in the limit $t\rightarrow \infty $,
the amplitude $\alpha $ in Eq. \eqref{Was} does not depend on the temperatures $T_1$ and $T_2$. To verify this rather strange prediction, we also depict by the
solid magenta curve in Fig. \ref{fig:omega} the result of a numerical
evaluation of the expression \eqref{omegapdf} for two quite
different temperatures $T_{1}=50$ and $T_{2}=1$ (note that other curves
correspond to $T_{1}=1$ and $T_{2}=4$) at time $t=30$. We observe that
even for this quite moderate value of $t$, the curve is very close to our
prediction in Eq. \eqref{zut} thereby confirming the temperature-independent
form of the long-time limit in Eq. \eqref{zut}.  While we are unable to provide simple physical arguments explaining this intriguing behavior, it seems to be quite evident from the mathematical point of view. Indeed, the angular velocity is formally defined as the ratio of the angular momentum and the moment of inertia ${I = x^2 + y^2}$ (see Eq. \eqref{meanomega}) which is formally equal to $2 U_{u=0}(t)/\lambda$, where $U_{u=0}(t)$ is the potential energy of the particle at zero value of the coupling parameter $u$. In the next Section, we determine the large-$t$ asymptotic form of the probability density  of  $U(t)$, see Eq. \eqref{energy}, and show that the temperatures enter this function only via their sum $T_1 + T_2$. In other words, for $t \to \infty$ and
 for most of realizations of the process we have $I = (T_1 + T_2) {\cal I}$, where ${\cal I}$ is a temperature-independent random variable with exponential distribution. In turn,  our analysis of the large-$t$ asymptotic form of the probability density
  of the angular momentum (see Eq. \eqref{PLas}) also shows that the latter incorporates the temperatures
 only via their sum, suggesting that for the most of the realizations the angular momentum behaves as $L = (T_1 + T_2) {\cal L}$ where ${\cal L}$ is temperature-independent. In view of this argument, it does not seem surprising that the large-$t$ and large-$W$ asymptotic form of the probability density function of $W = {\cal I}/{\cal L}$ is independent of the temperatures.

\section{Kinetic, potential and total energy}
\label{sec:energy}

The energy is continuously pumped into the system and, in absence of a dissipation,
is therefore increasing with time.
Concurrently, since the particle performs a random motion, its energy is a random variable and it seems interesting to study its statistical properties. We consider separately the potential energy $U(t)$, defined in Eq. \eqref{3}, kinetic energy
\begin{equation}
K(t) =  \left(\dot{x}^2 + \dot{y}^2\right)/2 \,,
\end{equation}
and the total energy
$E(t) = U(t) + K(t)$.

Using the
expressions for the moments of positions and velocities, derived in Sec. \ref{sec:3},
we readily find that
the first moments of the energies obey
\begin{equation}
\label{UKmean}
\begin{split}
\left \langle U(t) \right \rangle & = \frac{(T_{1}+T_{2})}{2}t-\frac{\left( T_{1}+T_{2}\right) }{8}\left( \frac{%
\sin \left( 2\Omega _{+}t\right) }{\Omega _{+}}+\frac{\sin \left( 2\Omega
_{-}t\right) }{\Omega _{-}}\right) \,, \\
\left \langle K(t) \right \rangle &  =\frac{(T_{1}+T_{2})}{2}t+\frac{\left( T_{1}+T_{2}\right) }{8}\left( \frac{%
\sin \left( 2\Omega _{+}t\right) }{\Omega _{+}}+\frac{\sin \left( 2\Omega
_{-}t\right) }{\Omega _{-}}\right) \,,
\end{split}%
\end{equation}%
and therefore,
\begin{equation}
\label{Emean}
\left \langle E(t) \right \rangle =(T_{1}+T_{2}) \, t \,.
\end{equation}
We note that the mean potential and the kinetic energies are both linearly growing with time, in the leading order, and also contain some sub-dominant oscillatory terms.
The prefactor in the dominant linear dependence on time is just the sum of the temperatures, as it should be, and is independent of  the strength $\lambda$ of the potential and of the coupling parameter $u$.  In turn, the mean total energy
is simply a  monotonically growing function of time, because the sub-dominant oscillatory terms cancel each other.

\begin{figure}[ht]
\includegraphics[width=0.60\textwidth,clip=]{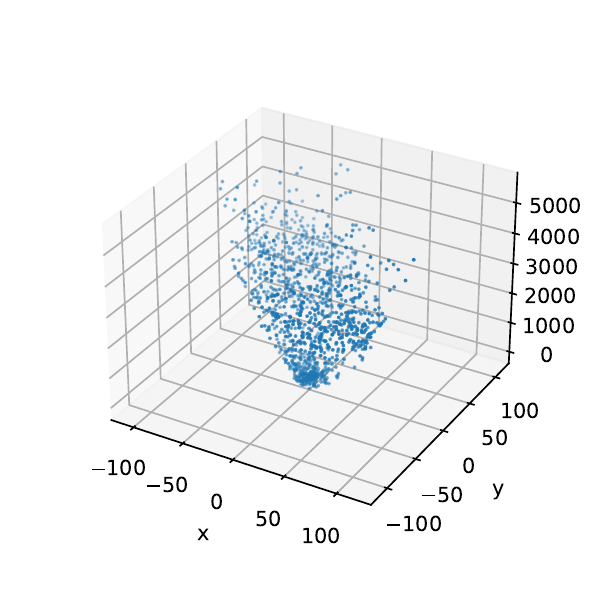}
\includegraphics[width=0.5\textwidth,clip=]{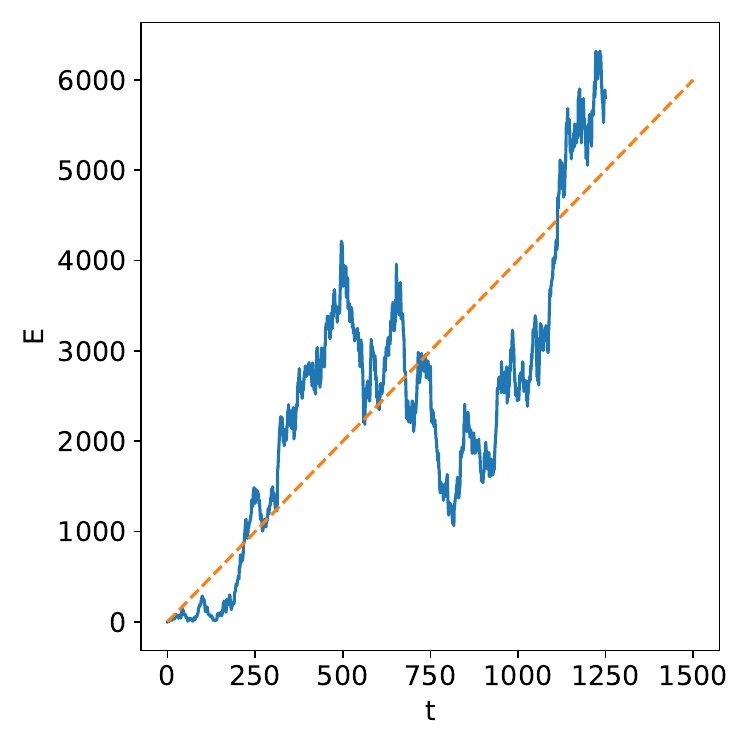}
\caption{ The total energy $E(t)$ of the particle for a given realization of noises, $u=1/2$, $T_1 = 1$ and $T_2 = 2$. Left panel: The spread of $E(t)$ for consecutive positions of the particle on the $(x,y)$-plane. Right panel : $E(t)$ as function of time for a given trajectory of the particle (blue noisy curve). The dashed line depicts the mean total energy $\langle E(t) \rangle$  in Eq. \eqref{Emean}. }\label{fig:traj}
\end{figure}

In Fig. \ref{fig:traj} we depict the total energy of the particle for a single realization
of the particle's trajectory.  We observe significant fluctuations: indeed,
the spread of realization-dependent values around the mean one appears to be very large. Therefore, to fully quantify the temporal evolution of the energies
it is necessary
to go beyond the mean  values and to determine the full probability density functions of the energies. This can be done by standard means, i.e., by evaluating first the respective characteristic functions and then inverting the corresponding Laplace transforms. In doing so and omitting the intermediate calculations, we find that the probability densities $P_t(U)$ and $P_t(K)$ of the potential and of the kinetic energies are
\begin{equation}
\begin{split}
\label{ener}
P_t(U) &=  \frac{1}{\lambda \sqrt{(1-u^2) d}} \exp\left(-\frac{\langle U \rangle}{(1 - u^2) d \lambda^2} U\right) I_0\left(b \, U\right),  \\
b &= \frac{\sqrt{\langle U\rangle^2 - (1- u^2) d \lambda^2}}{(1- u^2) d \lambda^2} \,, \quad d = \langle x^2 \rangle \langle y^2\rangle - \langle x y\rangle^2 \,,\\
P_t(K)  &= \frac{1}{\sqrt{d'}} \exp\left(- \frac{\langle K \rangle}{d'} K\right) I_{0}\left(\frac{\sqrt{\langle K \rangle^2 - d'}}{d'} K\right) \,, \quad
d' = \langle \dot{x}^2 \rangle \langle \dot{y}^2 \rangle - \langle \dot{x} \dot{y}\rangle^2
\end{split}
\end{equation}
where $I_0(z)$ is the modified Bessel function, while $\langle U \rangle$ and $\langle K\rangle$ are defined in Eqs. \eqref{UKmean}.  In turn, the probability density  of the total energy cannot be obtained in a closed form but rather in form of the inverse Laplace transform of the inverse of a square root of a fourth-order polynomial of the Laplace parameter $s$:
\begin{equation}
\label{total}
P_t(E) = {\cal L}^{-1}_{s,E} \left\{\frac{1}{\sqrt{\left(1 + 2 \langle K\rangle s + d' \, s^2\right) \left(1 + 2 \langle U \rangle s + (1 - u^2) \lambda^2 d \, s^2\right)}} \right\} \,.
\end{equation}
To quantify fluctuations of the energies, we determine their variances. This can be done directly by integrating the probability density functions in Eqs. \eqref{ener}, and by differentiating the kernel function in Eq. \eqref{total} with respect to $s$ and then setting $s=0$. This gives the following exact expressions
\begin{equation}
\begin{split}
\mathrm{Var}\{U\} &= 2 \langle U\rangle^2 - (1-u^2) d \lambda^2 \,, \\
\mathrm{Var}\{K\} &= 2 \langle K\rangle^2 - d'  \,,\\
\mathrm{Var}\{E\} &= 2 \langle U\rangle^2 + 2 \langle K\rangle^2 - d' -  (1-u^2) d \lambda^2 \,.
\end{split}
\end{equation}
 In particular, we find that the coefficients of variation of the probability density of the energies admits the limits
\begin{equation}
\lim_{t \to \infty} \frac{\sqrt{\mathrm{Var}\{U\} } }{\langle U \rangle } = \lim_{t \to \infty} \frac{\sqrt{\mathrm{Var}\{K\} } }{\langle K \rangle }  = \lim_{t \to \infty} \frac{\sqrt{\mathrm{Var}\{E\} } }{\langle E \rangle }  = 1 \,,
\end{equation}
 signifying that fluctuations are exactly of the same order of magnitude as the mean values themselves. Consequently, the corresponding probability densities are effectively broad (see, e.g., \cite{mejia}) and the mean values do not characterize the behavior of energies adequately well.

\begin{figure}[t!]
        \includegraphics[scale=0.35]{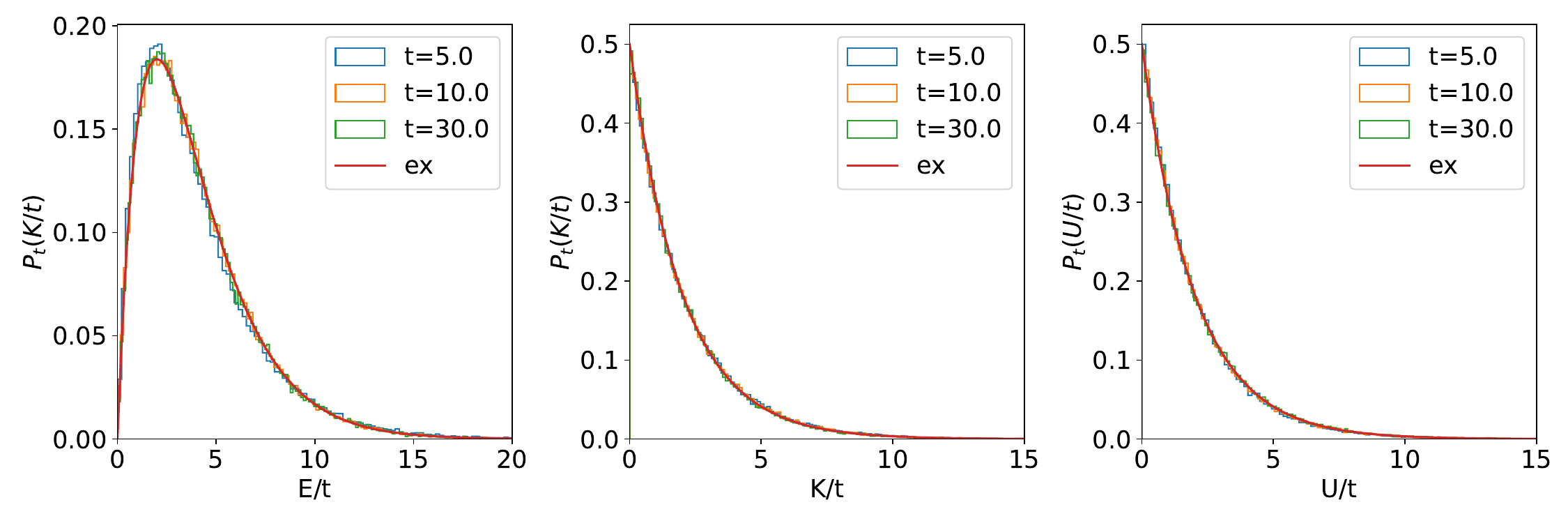}
        \caption{Probability densities of the energies. Solid (red) curves depict the asymptotic forms in Eq. \eqref{energy}, while the noisy curves represents the probability density functions deduced from numerical simulations.}
        \label{fig:E}
\end{figure}

Lastly, we analyze the large-energy tails of the probability densities
(\ref{ener}) and (\ref{total}). It is quite straightforward for the potential and kinetic energies whose distributions are given by simple explicit forms in Eqs. \eqref{ener} and only slightly more involved
for the total energy whose probability density  is given implicitly, in form of the inverse Laplace transform in Eq. \eqref{total}. We proceed here exactly in the same way as in Sec. \ref{sec:gyration} for the evaluation of the limiting form of the probability density of the angular moment. That being, we turn to the variable $E/t$ and consider the large-$t$ behavior of the coefficients of the fourth-order polynomial in $s$. In doing so, we eventually find that all three probability densities have simple exponential tails of the form :
\begin{equation}
\begin{split}
\label{energy}
P_{t \to \infty}(U) &\simeq  \frac{2}{(T_1+T_2)t}\exp\left( -\frac{2 U}{(T_1+T_2)t}\right)  \,,\\
P_{t\to \infty}(K) &\simeq \frac{2}{(T_1+T_2)t}\exp\left(- \frac{2 K}{(T_1+T_2)t}\right) \,, \\
P_{t \to \infty}(E)& \simeq \frac{4 E}{(T_1+T_2)^2 t^2}\exp\left(-\frac{2 E}{(T_1+T_2)t}\right) \,.
\end{split}
\end{equation}
The probability densities in Eqs. \eqref{energy} are depicted in Fig \ref{fig:E} together with the corresponding
forms obtained in the numerical simulations. We observe an excellent agreement between our theoretical   predictions and numerics even for quite modest times. Lastly, we note that, in contrast to the probability densities of the kinetic and the potential energies which are monotonically decreasing functions, the probability density function of the  total  energy has a maximum, which determines the most probable total energy, i.e., the value that should be observed for the majority of realizations of the process,
\begin{equation}
E_{\mathrm{mp}} = \frac{(T_1 + T_2) t}{2} \,,
\end{equation}
which appears to be two times smaller than the mean total energy, Eq. \eqref{Emean}. This again signifies
that fluctuations are very important.

\section{Conclusions}
\label{sec:conc}

 To conclude, we studied here
the dynamics of a particle which moves randomly on a plane and the position
components are defined as two linearly coupled random-acceleration processes
evolving in a parabolic confining potential.  Each position component is subject to its own independent Gaussian noise
with the amplitude (temperature) is, in general, not equal to that of the noise acting on the other component.
Our analysis was motivated, in part, by recent interesting observations made in \cite{Mancois2018,Bae2021} for
a finite-mass Brownian gyrator model in the non-equilibrium steady-state attained in the limit $t \to \infty$. Here we concentrated rather on the large-\_\ but finite-time behavior in a somewhat simplified system in which the damping (and hence, the dissipation)
 is set equal to zero. Therefore, apart of being of an interest in its own right, our model and the results can be considered as describing
 the temporal evolution of a finite-mass gyrator at transient stages, if the mass is sufficiently small.

 In addition to the standard characteristics, such as, e.g., the moments and the mixed moments, the two-time correlations
and the position-velocity probability density function, we also determined the characteristics of the rotational motion - the angular momentum and the angular velocity. We have shown that in case when the amplitudes  of noises acting on the components are not equal, the angular momentum and the angular velocity have non-zero mean values.
However, unlike it happens for the Brownian gyrator for which these properties approach constant values, for the model under study they show irregular oscillations with time, meaning that the torque exerted on the particle aperiodically changes its sign at different time moments and the particle is prompted to rotate clockwise and then counter-clockwise.

Looking on these random variables from a broader perspective, we found the full probability densities of the angular momentum $L$ and of the angular velocity $W$. We showed that the former has simple exponential large-$L$ tails for any large but fixed time, and hence, all moments are finite at a finite $t$. These asymptotic tails are symmetric with respect to the sign of $L$ and hence, a non-zero value of the first moment stems from the asymmetry of the probability density  present at small values of $L$.
In the limit $t \to \infty$, this latter probability density function converges to a uniform distribution with a diverging variance, which signifies that fluctuations become very significant and the angular momentum is not self-averaging. In turn, we realized that the probability density  of the angular velocity possesses heavy power-law tails  $1/|W|^3$ and hence, the mean angular velocity is the only existing moment. Therefore, fluctuations of the angular velocity destroy any systematic rotational motion.

We note parenthetically that the heavy tails of the form $1/|W|^3$ are exactly the same
that were previously found for the standard Brownian gyrator \cite{viot2023destructive}. Given that the dynamics in both models is quite different, one may  conclude that such tails is a generic feature resulted from the Gaussian noise acting on the particle. In this regard, it seems interesting to verify whether this feature will remain valid also for a more general model of a finite-mass Brownian gyrator. 

\medskip
\textbf{Acknowledgment} L.P. is grateful to the Ecole Normale Superieure (Paris, France) and the Institut des Hautes Etudes Scientifiques (Bures sur Yvette, France) for providing him a stay during which part of this work has been carried out.

\bibliographystyle{plain}


\begin{appendices}
\section{The proof of invertibility of $M(t)$}\label{secA0}

It is convenient to rewrite the system (\ref{2}) -- (\ref{initial}) of two
equation of second order for $x(t)$ and $y(t)$ as the system of four equation
of first order for $X(t)$ of (\ref{X}):%
\begin{equation}
\overset{.}{X}(t)=AX(t)+B\mathbf{\xi} (t),\;X(0)=0,  \label{XAB}
\end{equation}%
where
\begin{equation*}
A=\left(
\begin{array}{cc}
\mathbf{0} & \mathbf{1}_{} \\
-\mathbf{a} & \mathbf{0}%
\end{array}%
\right) ,\;B=\left(
\begin{array}{cc}
\mathbf{0} & \mathbf{0} \\
\mathbf{0} & \mathbf{1}%
\end{array}%
\right) ,\;\xi(t)=\left(
\begin{array}{c}
0 \\
\mathbf{\xi }^{(2)}(t)%
\end{array}%
\right) ,
\end{equation*}%
i.e., $A$ and $B$ are $4\times 4$ matrices written as $2\times 2$ matrices
with $2\times 2$ blocks and $\xi $ is $4\times 1$ vector written as $2\times
1$ vector with $2\times 1 $ components
\begin{equation*}
\mathbf{0}=\left(
\begin{array}{cc}
0 & 0 \\
0 & 0%
\end{array}%
\right) ,\;\mathbf{1}_{}=\left(
\begin{array}{cc}
1 & 0 \\
0 & 1%
\end{array}%
\right) ,\;\mathbf{a}=\lambda \left(
\begin{array}{cc}
1 & u \\
u & 1%
\end{array}%
\right) ,\;\mathbf{\xi}^{(2)}(t) =\left(
\begin{array}{c}
\xi _{1}(t) \\
\xi _{2}(t)%
\end{array}%
\right) .
\end{equation*}%
Then%
\[
X(t)=\int_{0}^{t}e^{A(t-s)}B\xi (s)ds,
\]%
and (\ref{noises})  and (\ref{eq:MatM}) yield%
\begin{equation}
M(t)=\int_{0}^{t}e^{As}\mathcal{B}e^{A^{T}s}ds,  \label{Cint}
\end{equation}%
where%
\begin{equation*}
\mathcal{B}=\left(
\begin{array}{cc}
\mathbf{0} & \mathbf{0} \\
\mathbf{0} & \mathbf{b}%
\end{array}%
\right) ,\;\mathbf{b}=\left(
\begin{array}{cc}
2T_{1}/m^{2} & 0 \\
0 & 2T_{2}/m^{2}%
\end{array}%
\right).
\end{equation*}
We have%
\begin{equation*}
A^{2}=-\left(
\begin{array}{cc}
\mathbf{a} & \mathbf{0} \\
\mathbf{0} & \mathbf{a}%
\end{array}%
\right) =-\left(
\begin{array}{cc}
\mathbf{a}^{1/2} & \mathbf{0} \\
\mathbf{0} & \mathbf{a}^{1/2}%
\end{array}%
\right) ^{2}=-\mathcal{A}^{2},
\end{equation*}%
hence,
$e^{As}=\cos \mathcal{A}s +  A(\sin \mathcal{A}s)/\mathcal{A}.
$ This, the spectral expansion of%
\begin{equation*}
\mathbf{a}=\lambda (1+u)|\psi _{+}\rangle \langle \psi _{+}|+\lambda
(1-u)|\psi _{-}\rangle \langle \psi _{-}|,\;|\psi _{+}\rangle
=2^{-1/2}\left(
\begin{array}{c}
1 \\
\pm 1%
\end{array}%
\right).
\end{equation*}%
and a simple but  somewhat tedious algebra allow us to show that
$M(t)$ is invertible for any $t>0$.

\section{Characteristic function of the angular  momentum}\label{secA1}

We start with the formal definition
\begin{equation}
\label{La}
\begin{split}
\Phi_{L}(\nu) &= \left \langle \exp\left(i \nu \left(x \dot{y} - y \dot{x}\right)\right) \right \rangle \\
& = \int^{\infty}_{-\infty} \int^{\infty}_{-\infty} \int^{\infty}_{-\infty} \int^{\infty}_{-\infty} dx \, dy \, d\dot{x} \, d\dot{y} \, \exp\left(i \nu \left(x \dot{y} - y \dot{x}\right)\right) \, {\cal P}_t\left(x,y,\dot{x},\dot{y}\right) \,.
\end{split}
\end{equation}
Expressing the joint position-velocity probability density through its characteristic function, we have
that the characteristic function of the angular momentum can be formally written as
\begin{equation}
\label{Lb}
\begin{split}
&\Phi_{L}(\nu) =  \frac{1}{(2 \pi)^4}  \int^{\infty}_{-\infty} \int^{\infty}_{-\infty} \int^{\infty}_{-\infty} \int^{\infty}_{-\infty} d\omega_1 \, d\omega_2 \, d\omega_3 \, d\omega_4 \, \Phi_t(\boldsymbol{\omega}) \\
&\times \int^{\infty}_{-\infty} \int^{\infty}_{-\infty} \int^{\infty}_{-\infty} \int^{\infty}_{-\infty} dx \, dy \, d\dot{x} \, d\dot{y} \, \exp\left(i \nu \left(x \dot{y} - y \dot{x}\right) - i \omega_1 x - i \omega_2 y - i \omega_3 \dot{x} - i \omega_4 \dot{y}\right) \\
&=  \frac{1}{(2 \pi)^2}  \int^{\infty}_{-\infty} \int^{\infty}_{-\infty} \int^{\infty}_{-\infty} \int^{\infty}_{-\infty} d\omega_1 \, d\omega_2 \, d\omega_3 \, d\omega_4 \, \Phi_t(\boldsymbol{\omega}) \\
& \times \int^{\infty}_{-\infty} \int^{\infty}_{-\infty}  d\dot{x} \, d\dot{y} \, \exp\left(- i \omega_3 \dot{x} - i \omega_4 \dot{y}\right) \delta\left(\nu \dot{y} - \omega_1\right) \, \delta\left(\nu \dot{x} + \omega_2\right) \,.
\end{split}
\end{equation}
Performing next the integrals over $\dot{x}$ and $\dot{y}$, we get
\begin{equation}
\begin{split}
\label{Lc}
\Phi_{L}(\nu)& =  \frac{1}{(2 \pi)^2 \nu ^2}  \int^{\infty}_{-\infty} \int^{\infty}_{-\infty} \int^{\infty}_{-\infty} \int^{\infty}_{-\infty} d\omega_1 \, d\omega_2 \, d\omega_3 \, d\omega_4 \, \Phi_t(\boldsymbol{\omega}) \exp\left(i \frac{\omega_2 \omega_3}{\nu} - i \frac{\omega_1 \omega_4}{\nu}\right) \,.
\end{split}
\end{equation}
Performing Gaussian integrals yields the exact expression in Eqs. \eqref{La1} and \eqref{eq:Mat}.

\section{Characteristic function of the angular velocity}\label{secA2}

We start with the formal definition
\begin{equation}
\label{omegaa}
\begin{split}
\Phi_{W}(z) &= \left \langle \exp\left(i z \frac{\left(x \dot{y} - y \dot{x}\right)}{x^2 + y^2}\right) \right \rangle \\
& = \int^{\infty}_{-\infty} \int^{\infty}_{-\infty} \int^{\infty}_{-\infty} \int^{\infty}_{-\infty} dx \, dy \, d\dot{x} \, d\dot{y} \, \exp\left(i z \frac{\left(x \dot{y} - y \dot{x}\right)}{x^2 + y^2}\right) \, {\cal P}_t\left(x,y,\dot{x},\dot{y}\right) \,.
\end{split}
\end{equation}
Expressing again the joint position-velocity probability density through its characteristic function, we find
that the characteristic function of the angular velocity can be formally written as
\begin{equation}
\label{omegab}
\begin{split}
&\Phi_{W}(z) =  \frac{1}{(2 \pi)^4}  \int^{\infty}_{-\infty} \int^{\infty}_{-\infty} \int^{\infty}_{-\infty} \int^{\infty}_{-\infty} d\omega_1 \, d\omega_2 \, d\omega_3 \, d\omega_4 \, \Phi_t(\boldsymbol{\omega}) \\
&\times \int^{\infty}_{-\infty} \int^{\infty}_{-\infty} \int^{\infty}_{-\infty} \int^{\infty}_{-\infty} dx \, dy \, d\dot{x} \, d\dot{y} \, \exp\left(i z \frac{\left(x \dot{y} - y \dot{x}\right)}{x^2 + y^2} - i \omega_1 x - i \omega_2 y - i \omega_3 \dot{x} - i \omega_4 \dot{y}\right) \\
&=  \frac{1}{(2 \pi)^2}  \int^{\infty}_{-\infty} \int^{\infty}_{-\infty} \int^{\infty}_{-\infty} \int^{\infty}_{-\infty} d\omega_1 \, d\omega_2 \, d\omega_3 \, d\omega_4 \, \Phi_t(\boldsymbol{\omega}) \\
& \times \int^{\infty}_{-\infty} \int^{\infty}_{-\infty}  dx \, dy \, \exp\left(- i \omega_1 x - i \omega_2 y\right) \delta\left(\frac{z y}{x^2+y^2} + \omega_3\right) \, \delta\left(\frac{z x}{x^2+y^2} - \omega_4\right) \,.
\end{split}
\end{equation}
Performing first the integrals over $\omega_3$ and $\omega_4$, then over $\omega_1$ and $\omega_2$, we change
the integration variables $x$ and $y$ for polar coordinates  $x = \rho \cos(\theta)$ and $y = \rho \sin(\theta)$ to get
\begin{equation}
\begin{split}
\label{q}
&\Phi_{W}(z) = \frac{1}{2 \pi \sqrt{\langle x^2\rangle \langle y^2\rangle - \langle xy\rangle^2}} \int^{2 \pi}_0 d\theta  \int^{\infty}_0 \rho \, d\rho \, \exp\Big(- \frac{A_{\theta} \rho^2  + i B_{\theta} z + C_{\theta} z^2/\rho^2}{2 \left(\langle x^2\rangle \langle y^2\rangle - \langle xy\rangle^2\right)} \Big) \,,
\end{split}
\end{equation}
where the functions $A_{\theta}$, $B_{\theta}$ and $C_{\theta}$ are given explicitly by
\begin{equation}
\begin{split}
\label{cts}
A_{\theta} & =\langle x^2\rangle \sin^2(\theta) + \langle y^2\rangle \cos^2(\theta) - \langle xy\rangle \sin(2 \theta)\\
B_{\theta} &= 2 \left(\langle xy \rangle \langle  y \dot{y } \rangle  - \langle y^2\rangle \langle x\dot{y}\rangle\right) \cos^2(\theta) - 2 \left(\langle xy\rangle \langle x\dot{x}\rangle  - \langle x^2\rangle \langle y\dot{x} \rangle\right) \sin^2(\theta) \\
&+ \left(\langle xy\rangle \langle x\dot{y}\rangle - \langle xy\rangle \langle y\dot{x} \rangle + \langle y^2\rangle \langle x\dot{x} \rangle - \langle x^2\rangle \langle y\dot{y} \rangle \right) \sin(2 \theta)\\
C_{\theta} &= \frac{1}{2} \Big[\left(\langle x^2\rangle \langle y^2\rangle - \langle xy\rangle^2\right) \left(\langle \dot{x}^2 \rangle + \langle \dot{y}^2 \rangle\right) - 2 \Big(\langle y^2\rangle \langle x\dot{y}\rangle^2 + \langle x^2\rangle \langle y \dot{y}\rangle^2 \\&- 2 \langle xy\rangle \langle x\dot{y} \rangle \langle y\dot{y}\rangle\Big) \cos^2(\theta)
- 2 \Big(\langle x^2\rangle \langle y\dot{x}\rangle^2 + \langle y^2\rangle \langle x \dot{x}\rangle^2 - 2 \langle xy\rangle \langle x\dot{x} \rangle \langle y\dot{x}\rangle\Big) \sin^2(\theta) \\&- \left(\langle x^2\rangle \langle y^2\rangle - \langle xy\rangle^2\right) \left(\langle \dot{x}^2 \rangle - \langle \dot{y}^2 \rangle\right) \cos(2 \theta)
+ 2 \Big(\langle xy \rangle^2 \langle \dot{x}\dot{y} \rangle + \langle x^2\rangle \langle y\dot{x} \rangle \langle y\dot{y} \rangle \\&+
 \langle y^2\rangle
\langle x\dot{x} \rangle  \langle x\dot{y} \rangle - \langle xy\rangle \langle x\dot{y}\rangle \langle y\dot{x}\rangle -
\langle xy\rangle \langle x\dot{x}\rangle \langle y\dot{y}\rangle - \langle x^2\rangle \langle y^2\rangle \langle \dot{x} \dot{y}  \rangle
\Big) \sin(2 \theta)
\Big] \,.
\end{split}
\end{equation}
Performing in Eq. \eqref{q}  the integration over $\rho$, we arrive at our Eq. \eqref{omegax}.

\end{appendices}

\end{document}